\documentstyle[preprint,eqsecnum,amsfonts,aps]{revtex}

\begin{document}
\draft
\title{Post-Newtonian Gravitational Radiation}
\author{Luc Blanchet}
\address{D\'epartement d'Astrophysique Relativiste et de Cosmologie,\\
Centre National de la Recherche Scientifique (UMR 8629),\\
Observatoire de Paris, 92195 Meudon Cedex, France}

\date{\today}
\maketitle

\def\Box{\hbox{\rlap{$\sqcup$}$\sqcap$}}

\section{Introduction}

\subsection{On approximation methods in general relativity}

Let us declare that the most important {\it devoir} of any physical
theory is to draw firm predictions for the outcome of laboratory
experiments and astronomical observations.  Unfortunately, the devoir
is quite difficult to fulfill in the case of general relativity,
essentially because of the complexity of the Einstein field equations,
to which only few exact solutions are known. For instance, it is
impossible to settle the exact prediction of this theory when there
are no symmetry in the problem (as is the case in the problem of the
gravitational dynamics of separated bodies). Therefore, one is often
obliged, in general relativity, to resort to approximation methods.

It is beyond question that approximation methods do work in general
relativity. Some of the great successes of this theory were in fact
obtained using approximation methods. We have particularly in mind the
test by Taylor and collaborators \cite{TFMc79,TW82,T93} regarding the
orbital decay of the binary pulsar PSR 1913+16, which is in agreement
to within 0.35\% with the general-relativistic {\it post-Newtonian}
prediction. However, a generic problem with approximation methods
(especially in general relativity) is that it is non trivial to define
a clear framework within which the approximation method is
mathematically well-defined, and such that the results of successive
approximations could be considered as {\it theorems} following some
precise (physical and/or technical) assumptions. Even more difficult
is the problem of the relation between the approximation method and
the {\it exact} theory. In this context one can ask: What is the
mathematical nature of the approximation series (convergent,
asymptotic, $\dots$)? What its ``reliability'' is (i.e., does the
approximation series come from the Taylor expansion of a family of
exact solutions)? Does the approximate solution satisfy some ``exact''
boundary conditions (for instance the no-incoming radiation
condition)?

Since the problem of theoretical prediction in general relativity is
complex, let us distinguish several approaches (and ways of thinking)
to it, and illustrate them with the example of the prediction for the
binary pulsar. First we may consider what could be called the
``physical'' approach, in which one analyses the relative importance
of each physical phenomena at work by using crude numerical estimates,
and where one uses only the lowest-order approximation, relating if
necessary the local physical quantities to observables by means of
balance equations (perhaps not well defined in terms of basic
theoretical concepts).  The physical approach to the problem of the
binary pulsar is well illustrated by Thorne in his beautiful Les
Houches review \cite{Th83} (see also the round table discussion
moderated by Ashtekar \cite{A83}): one derives the loss of energy by
gravitational radiation from the (Newtonian) quadrupole formula
applied formally to point-particles, assumed to be test-masses though
they are really self-gravitating, and one argues ``physically'' that
the effect comes from the variation of the Newtonian binding energy in
the center-of-mass frame -- indeed, on physical grounds, what else
could this be (since we expect the rest masses won't vary)? The
physical approach yields the correct result for the rate of decrease
of the period of the binary pulsar. Of course, thinking physically is
extremely useful, and indispensable in a preliminary stage, but
certainly it should be completed by a solid study of the connection to
the mathematical structure of the theory. Such a study would {\it a
posteriori} demote the physical approach to the status of
``heuristic'' approach. On the other hand, the physical approach may
fall short in some situations requiring a sophisticated mathematical
modelling (like in the problem of the dynamics of singularities),
where one is often obliged to follow one's mathematical rather than
physical insight.

A second approach, that we shall qualify as ``rigorous'', has been
advocated mainly by J\"urgen Ehlers (see, e.g., \cite{Ehlfolk}). It
consists of looking for a high level of mathematical rigor, within the
exact theory if possible, and otherwise using an approximation scheme
that we shall be able to relate to the exact theory. This does not
mean that we will be so much wrapped up by mathematical rigor as to
forget about physics. Simply, in the rigorous approach, the prediction
for the outcome of an experiment should follow mathematically from
first theoretical principles. Clearly this approach is the one we
should ideally adhere to. As an example, within the rigorous approach,
one was not permitted, by the end of the seventies, to apply the
standard quadrupole formula to the binary pulsar. Indeed, as pointed
out by Ehlers {\it et al} \cite{EhlRGH}, it was not clear that
gravitational radiation reaction on a self-gravitating system implies
the standard quadrupole formula for the energy flux, notably because
computing the radiation reaction demands {\it a priori} three
non-linear iterations of the field equations \cite{WalW80}, which were
not fully available at that time. Ehlers and collaborators
\cite{EhlRGH} remarked also that the exact results concerning the
structure of the field at infinity (notably the asymptotic shear of
null geodesics whose variation determines the flux of radiation) were
not connected to the actual dynamics of the binary.

Maybe the most notable result of the rigorous approach concerns the
relation between the exact theory and the approximation methods. In
the case of the post-Newtonian approximation (limit $c\to\infty$),
J\"urgen Ehlers has provided with his frame theory
\cite{Ehl91,Ehl97,Ehl98} a conceptual framework in which the
post-Newtonian approximation can be clearly formulated (among other
purposes). This theory unifies the theories of Newton and Einstein
into a single generally covariant theory, with a parameter $1/c$
taking the value zero in the case of Newton and being the inverse of
the speed of light in the case of Einstein. Within the frame theory
not only does one understand the limit relation of Einstein's theory
to Newton's, but one explains why it is legitimate when describing the
predictions of general relativity to use the common-sense language of
Newton (for instance thinking that the trajectories of particles in an
appropriately defined coordinate system take place in some Euclidean
space, and viewing the coordinate velocities as being defined with
respect to absolute time). It was shown by Lottermoser \cite{Lott90}
that the constraint equations of the (Hamiltonian formulation of the)
Ehlers frame theory admit solutions with a well-defined post-Newtonian
limit. Further in the spirit of the rigorous approach, we quote the
work of Rendall \cite{Rend92} on the definition of the post-Newtonian
approximation, and the link to the post-Newtonian equations used in
practical computations.  (See also \cite{Fu83,FuSc83} for an attempt
at showing, using restrictive assumptions, that the post-Newtonian
series is asymptotic.)

The important remarks of J\"urgen Ehlers {\it et al} \cite{EhlRGH} on
the applicability of the quadrupole formula to the binary pulsar
stimulated research to settle down this question with (al least)
acceptable mathematical rigor. The question was finally answered
positively by Damour and collaborators \cite{BeDD81,DD81a,D82,D83a},
who obtained in algebraically closed form the general-relativistic
equations of motion of two compact objects, up to the requisite 5/2
post-Newtonian order (2.5PN order or $1/c^5$) where the gravitational
radiation reaction force appears. This extended to 2.5PN order the
work at 1PN of Lorentz and Droste \cite{LD17}, and Einstein, Infeld
and Hoffmann \cite{EIH}.  The net result is that the dynamics of the
binary pulsar as predicted by (post-Newtonian) general relativity is
in full agreement both with the prediction of the quadrupole formula,
as derived earlier within the ``physical'' approach, and with the
observations by Taylor {\it et al} (see \cite{D83b} for discussion).

Motivated by the success of the theoretical prediction in the case of
the binary pulsar \cite{BeDD81,DD81a,D82,D83a,D83b}, we shall try to
follow in this article the spirit of the ``rigorous'' approach of
J\"urgen Ehlers, notably in the way it emphasizes the mathematical
proof, but we shall also differ from it by a systematic use of
approximation methods. This slightly different approach recognizes
from the start that in certain difficult problems, it is impossible to
derive a physical result all the way through the exact theory without
any gap, so that one must proceed with approximations. {\it But}, in
this approach, one implements a mathematically well-defined framework
for the approximation method, and within this framework one proves
theorems that (ideally) guarantee the correctness of the theoretical
prediction to be compared with experiments. Because the comparison
with experiments is the only thing which matters {\it in fine} for a
pragmatist, we qualify this third approach as ``pragmatic''.

In this article we describe the pragmatic approach to the problem of
gravitational radiation emitted by a general isolated source, based on
the rigorous post-Minkowskian iteration of the field outside the
source \cite{BD86}, and on the general connection of the exterior
field to the field inside a slowly-moving source
\cite{B95,B98mult}. Note that for this particular problem the
pragmatic approach is akin to the rigorous one in that it permits to
establish some results on the connection between approximate and exact
methods. For instance it was proved by Damour and Schmidt
\cite{DSch90} (see also \cite{CSch79,Rend90}) that the
post-Minkowskian algorithm generates an asymptotic approximation to
exact solutions, and it was shown \cite{B87} that the solution
satisfies to any order in the post-Minkowskian expansion a rigorous
definition of asymptotic flatness at future null infinity. However it
remains a challenge to analyse in the manner of the rigorous approach
the relation to exact theory of the whole formalism of
\cite{BD86,B95,B98mult,B87}.

By combining the latter post-Minkowskian approximation and a
post-Newtonian expansion inside the system, it was proved (within this
framework of approximations) that the quadrupole formula for
slowly-moving, weakly-stressed and self-gravitating systems is
correct, even including post-Newtonian corrections \cite{BD89}; and
{\it idem} for the radiation reaction forces acting locally inside the
system, and for the associated balance equations \cite{B93,B97}. These
results answered positively Ehlers' remarks \cite{EhlRGH} in the case
of slowly-moving extended (fluid) systems. However we are also
interested in this article to the application to binary systems of
compact objects modelled by point-masses. Indeed the latter sources of
radiation are likely to be detected by future gravitational-wave
experiments, and thus concern the pragmatist. We shall see how one can
address the problem in this case. (When specialized to point-mass
binaries, the results on radiation reaction \cite{B93,B97} are in
agreement with separate work of Iyer and Will \cite{IW93,IW95}.) For
other articles on the problem of gravitational radiation from general
and binary point-mass sources, see \cite{D86,D300,Th300,W94,Bhouches}.

\subsection{Field equations and the no-incoming radiation condition}

The problem is to find the solutions, in the form of analytic
approximations, of the Einstein field equations in $I\!\!R^4$,

\begin{equation}
R^{\mu\nu}-{1\over 2}g^{\mu\nu}R={8\pi G\over c^4}T^{\mu\nu}\ ,
\end{equation}
and thus also of their consequence, the equations of motion of the
matter source, $\nabla_\nu T^{\mu\nu} =0$.  Throughout this work we
assume the existence and unicity of a global harmonic (or de Donder)
coordinate system. This means that we can choose the gauge condition

\begin{equation}
\partial_\nu h^{\mu\nu} = 0~;\qquad h^{\mu\nu}\equiv \sqrt{-g} g^{\mu\nu}-
\eta^{\mu\nu} \ ,
\end{equation}
where $g$ and $g^{\mu\nu}$ denote the determinant and inverse of the
covariant metric $g_{\mu\nu}$, and where $\eta^{\mu\nu}$ is an
auxiliary flat metric [i.e. $\eta^{\mu\nu}={\rm
diag}(-1,1,1,1)=\eta_{\mu\nu}$].  The Einstein field equations (1) can
then be replaced by the {\it relaxed} equations

\begin{equation}
\Box h^{\mu\nu} = {16\pi G\over c^4} \tau^{\mu\nu}\ ,
\end{equation} 
where the box operator is the flat d'Alembertian,
$\Box\equiv\Box_\eta=\eta^{\mu\nu}\partial_\mu\partial_\nu$, and where
the source term is the sum of a matter part and a gravitational part,

\begin{equation}
\tau^{\mu\nu} \equiv |g| T^{\mu\nu} + {c^4 \over 16\pi G}
\Lambda^{\mu\nu}\ .
\end{equation} 
In harmonic coordinates the field equations take the form of simple
wave equations, but whose source term is actually a complicated
functional of the gravitational field $h^{\mu\nu}$; notably the
gravitational part depends on $h^{\mu\nu}$ and its first and second
space-time derivatives:

\begin{eqnarray}
\Lambda^{\mu\nu} = &-& h^{\rho\sigma}
\partial^2_{\rho\sigma} h^{\mu\nu}+\partial_\rho h^{\mu\sigma} 
\partial_\sigma h^{\nu\rho} 
+{1\over 2}g^{\mu\nu}g_{\rho\sigma}\partial_\lambda h^{\rho\tau}
\partial_\tau h^{\sigma\lambda} \nonumber\\
&-&g^{\mu\rho}g_{\sigma\tau}\partial_\lambda h^{\nu\tau} 
\partial_\rho h^{\sigma\lambda} 
-g^{\nu\rho}g_{\sigma\tau}\partial_\lambda h^{\mu\tau} 
\partial_\rho h^{\sigma\lambda} 
+g_{\rho\sigma}g^{\lambda\tau}\partial_\lambda h^{\mu\rho} 
\partial_\tau h^{\nu\sigma}\nonumber\\
&+&{1\over 8}(2g^{\mu\rho}g^{\nu\sigma}-g^{\mu\nu}g^{\rho\sigma})
(2g_{\lambda\tau}g_{\epsilon\pi}-g_{\tau\epsilon}g_{\lambda\pi})
\partial_\rho h^{\lambda\pi} 
\partial_\sigma h^{\tau\epsilon}\ .
\end{eqnarray}
The point is that $\Lambda^{\mu\nu}$ is at least quadratic in $h$, so
the relaxed field equations (3) are very naturally amenable to a
perturbative non-linear expansion. As an immediate consequence of the
gauge condition (2), the right side of the relaxed equations is
conserved in the usual sense, and this is equivalent to the equations
of motion of matter:

\begin{equation}
\partial_\nu\tau^{\mu\nu}=0\quad\Leftrightarrow\quad\nabla_\nu 
T^{\mu\nu} =0 \ .
\end{equation}
We refer to $\tau^{\mu\nu}$ as the total stress-energy pseudo-tensor
of the matter and gravitational fields in harmonic coordinates. Since
the harmonic coordinate condition is Lorentz covariant,
$\tau^{\mu\nu}$ is a tensor with respect to Lorentz transformations
(but of course not with respect to general diffeomorphisms).

In order to select the physically sensible solution of the field
equations in the case of a bounded system, one must choose some
boundary conditions at infinity, i.e. the famous no-incoming radiation
condition, which ensures that the system is truly isolated (no
radiating sources located at infinity). In principle the no-incoming
radiation condition is to be formulated at past null infinity ${\cal
J}^-$. Here, we shall simplify the formulation by taking advantage of
the presence of the Minkowski background $\eta_{\mu\nu}$ to define the
no-incoming radiation condition with respect to the Minkowskian past
null infinity ${\cal J}_{\rm M}^-$. Of course, this does not make
sense in the exact theory where only exists the metric $g_{\mu\nu}$
and where the metric $\eta_{\mu\nu}$ is fictituous, but within
approximate (post-Minkowskian) methods it is legitimate to view the
gravitational field as propagating on the flat background
$\eta_{\mu\nu}$, since $\eta_{\mu\nu}$ does exist at any finite order
of approximation.

We formulate the no-incoming radiation condition in such a way that it
suppresses any homogeneous, regular in $I\!\!R^4$, solution of the
d'Alembertian equation $\Box h=0$. We have at our disposal the
Kirchhoff formula which expresses $h({\bf x}',t')$ in terms of values
of $h({\bf x},t)$ and its derivatives on a sphere centered on ${\bf
x}'$ with radius $\rho\equiv |{\bf x}'-{\bf x}|$ and at retarded time
$t\equiv t'-\rho/c$:

\begin{equation}
h({\bf x}',t')=\int\!\!\int {d\Omega\over 4\pi}\biggl[{\partial\over
\partial \rho}(\rho h) +{1\over c}{\partial\over\partial t}(\rho
h)\biggr]({\bf x},t)
\end{equation}
where $d\Omega$ is the solid angle spanned by the unit direction
$({\bf x}-{\bf x}')/\rho$. From the Kirchhoff formula we obtain the
no-incoming radiation condition as a limit at ${\cal J}_{\rm M}^-$,
that is $r\to +\infty$ with $t+r/c=$const (where $r=|{\bf x}|$).  In
fact we obtain two conditions: the main one,

\begin{equation}
\lim_{r\to +\infty\atop t+r/c={\rm const}}
\biggl[{\partial\over \partial r}(rh^{\mu\nu})
+{1\over c}{\partial\over\partial t}(rh^{\mu\nu})\biggr]({\bf x},t)=0\
,
\end{equation}
and an auxiliary condition, that $r\partial_\lambda h^{\mu\nu}$ should
be bounded at ${\cal J}_{\rm M}^-$, coming from the fact that $\rho$
in the Kirchhoff formula (7) differs from $r$ [we have $\rho=r-{\bf
x}'.{\bf n}+O(1/r)$ where ${\bf n}={\bf x}/r$].

In fact, we adopt in this article a much more restrictive condition of
no-incoming radiation, namely that the field is stationary before some
finite instant $-{\cal T}$ in the past:

\begin{equation}
t\leq -{\cal T} \,\Rightarrow\, {\partial \over \partial
t}[h^{\mu\nu}({\bf x},t)]=0 \ .
\end{equation}
In addition we assume that before $-{\cal T}$ the field
$h^{\mu\nu}({\bf x})$ is of order $O(1/r)$ when $r\to +\infty$. These
restrictive conditions are imposed for technical reasons following
\cite{BD86}, since they allow constructing rigorously (and proving
theorems about) the metric outside some time-like world tube $r\equiv
|{\bf x}|>{\cal R}$. We shall assume that the region $r>{\cal R}$
represents the exterior of an actual compact-support system with
constant radius $d <{\cal R}$ [i.e. $d$ is the maximal radius of the
adherence of the compact support of $T^{\mu\nu}({\bf x},t)$, for any
time $t$].

Now if $h^{\mu\nu}$ satisfies for instance (9), so does the
pseudo-tensor $\tau^{\mu\nu}$ built on it, and then it is clear that
the retarded integral of $\tau^{\mu\nu}$ satisfies itself the same
condition. Therefore one infers that the unique solution of the
Einstein equation (3) satisfying the condition (9) is

\begin{equation}
 h^{\mu\nu} = {16\pi G\over c^4} \Box ^{-1}_{R} \tau^{\mu\nu}\ ,
\end{equation}
where the retarded integral takes the standard form

\begin{equation}
(\Box ^{-1}_{R} \tau)({\bf x},t)\equiv-{1\over 4\pi}\int {d^3{\bf
x}'\over |{\bf x}-{\bf x}'|}\tau\left({\bf x}',t-|{\bf x}-{\bf
x}'|/c\right)\ .
\end{equation}
Notice that since $\tau^{\mu\nu}$ depends on $h$ and its derivatives,
the equation (10) is to be viewed rather as an integro-differential
equation equivalent to the Einstein equation (3) with no-incoming
radiation.

\subsection{Method and general physical picture}

We want to describe an isolated system, for instance a ``two-body
system'', in Einstein's theory. We expect (though this is not proved)
that initial data sets $g_{\mu\nu}$, $\partial_tg_{\mu\nu}$, $\rho$,
{\bf v} satisfying the constraint equations on the space-like
hypersurface $t=t_0$ exist, and that this determines a unique solution
of the field equations for any time $t$, which approaches in the case
of two bodies a ``scattering state'' when $t\to -\infty$, in which the
bodies move on unbound (hyperbolic-like) orbits. We assume that the
space-times generated by such data admit a past null infinity $\cal
J^-$ (or, if one uses approximate methods, $\cal J_{\rm M}^-$) with no
incoming radiation.  (Note that in a situation with initial scattering
the field might not satisfy the rigorous definitions of asymptotic
flatness at $\cal J^-$; see \cite{BaPr73,SSt79,WalW79,PSt81}.)  The
point to make is that in this class of space-times there is no degree
of freedom for the gravitational field (we could consider other
situations where the motion is influenced by incoming radiation).

Both our technical assumptions of compact support for the matter
source (with constant radius $d$) and stationarity before the time
$-{\cal T}$ contradict our expectation that a two-body system follows
an unbound orbit in the remote past. We do not solve this conflict but
argue as follows: (i) these technical assumptions permit to derive
rigorously some results, for instance the expression [given by (52)
with (56) below] of the far-field of an isolated past-stationary
system; (ii) it is clear that these results do not depend on the
constant radius $d$, and furthermore we check that they admit in the
``scattering'' situation a well-defined limit when $-{\cal T}\to
-\infty$ ; (iii) this makes us confident that the results are actually
valid for a more realistic class of physical systems which become
unbound in the past and are never stationary (and, even, one can give
{\it a posteriori} conditions under which the limit $-{\cal T}\to
-\infty$ exists for a general system at some order of approximation).

Suppose that the system is ``slowly-moving'' [in the sense of (12)
below], so that we can compute the field inside its compact support by
means of a post-Newtonian method, say $h_{\rm in}^{\mu\nu}\equiv
{\overline h}^{\mu\nu}$ where the overbar refers to the {\it formal}
post-Newtonian series. The post-Newtonian iteration (say, for
hydrodynamics) is not yet defined to all orders in $1/c$, but many
terms are known: see the works of Lorentz and Droste \cite{LD17},
Einstein, Infeld and Hoffmann \cite{EIH}, Fock \cite{Fock},
Chandrasekhar and collaborators \cite{C65,CN69,CE70}, Ehlers and
followers \cite{Ehl77,Ehl80,Ker80,Ker80',Capo81,BRu81,BRu82}, and many
other authors \cite{AD75,PapaL81,S85,S86,BD89,B95}.

On the other hand, outside the isolated system, the field is weak
everywhere and it satisfies the vacuum equations. Therefore, the
equations can be solved conjointly by means of a weak-field or
post-Minkowskian expansion ($G\to 0$), and, for each coefficient of
$G^n$ in the latter expansion, by means of a multipole expansion
(valid because we are outside). The general
Multipolar-post-Minkowskian (MPM) metric was constructed in
\cite{BD86,B87} as a functional of two sets of ``multipole moments''
$M_L(t)$ and $S_L(t)$ which were left arbitrary at this stage
(i.e. not connected to the source). The idea of combining the
post-Minkowskian and multipole expansions comes from the works of
Bonnor \cite{Bo59} and Thorne \cite{Th80}.  We denote by $h_{\rm
ext}^{\mu\nu}\equiv {\cal M}(h^{\mu\nu})$ the exterior solution, where
${\cal M}$ stands for the multipole expansion (as it will turn out,
the post-Minkowskian expansion appears in this formalism to be
somewhat less fundamental than the multipole expansion).

The key assumption is that the two expansions $h_{\rm
in}^{\mu\nu}={\overline h}^{\mu\nu}$ and $h_{\rm ext}^{\mu\nu}={\cal
M}(h^{\mu\nu})$ should match in a region of common validity for both
the post-Newtonian and multipole expansions. Here is where our
physical restriction to slow motion plays a crucial role, because such
an overlap region exists (this is the so-called exterior near-zone) if
and only if the system is slowly-moving. The matching is a variant of
the well-known method of matching of asymptotic expansions, very
useful in gravitational radiation theory
\cite{BuTh70,Bu71,K80,K80',AKKM82,BD89,DI91a,B95}.  It consists of
decomposing the inner solution into multipole moments (valid in the
outside), re-expanding the exterior solution in the near zone ($r/c\to
0$), and equating term by term the two resulting expansion
series. From the requirement of matching we obtain in \cite{B98mult},
and review in Sections 2 and 3 below, the general formula for the
multipole expansion ${\cal M}(h^{\mu\nu})$ in terms of the ``source''
multipole moments (notably a mass-type moment $I_L$ and a current-type
$J_L$), given as functionals of the {\it post-Newtonian} expansion of
the pseudo-tensor, i.e. ${\overline \tau}^{\mu\nu}$. [The previous
moments $M_L$ and $S_L$ (referred below to as ``canonical'') are
deduced from the source moments after a suitable coordinate
transformation.] In addition the matching equation determines the
radiation reaction contributions in the inner post-Newtonian metric
\cite{BD84,BD88,B97}.
 
To obtain the source multipole moments in terms of basic source
para\-meters (mass density, pressure), it remains to replace
${\overline \tau}^{\mu\nu}$ by the result of an explicit
post-Newtonian iteration of the inner field. This was done to 1PN
order in \cite{BD89,DI91a}, then to 2PN order in \cite{B95}, and the
general formulas obtained in \cite{B98mult} permit recovering these
results. See Section 6. On the other hand, if one needs the equations
of motion of the source, simply one inserts the post-Newtonian metric
into the conservation law $\partial_\nu{\overline
\tau}^{\mu\nu}=0$. (Note that we are speaking of the equations of
motion, which take for instance the form of Euler-type equations with
many relativistic corrections, but not of the {\it solutions} of these
equations, which are typically impossible to obtain analytically.)

From the harmonic coordinates, one can perform to all post-Minkowskian
orders \cite{B87} a coordinate transformation to some radiative
coordinates such that the metric admits a far-field expansion in
powers of the inverse of the distance $R$ (without the powers of $\ln
R$ which plague the harmonic coordinates). Considering the leading
order $1/R$ one compares the exterior metric, which is parametrized by
the source moments (connected to the source via the matching
equation), to the metric defined with ``radiative'' multipole moments,
say $U_L$ and $V_L$. This gives $U_L$ and $V_L$ in terms of the source
moments, notably $I_L$ and $J_L$, and {\it a fortiori} of the source
parameters. This solves, within approximate methods, the problem of
the relation between the far field and the source.  The radiative
moments have been obtained with increasing precision reaching now 3PN
\cite{BD92,B98quad,B98tail}, as reviewed in Section 5.

The previous scheme is developed for a general description of matter,
however restricted to be smooth (we have in mind a general
``hydrodynamical'' $T^{\mu\nu}$). Thus the scheme {\it a priori}
excludes the presence of singularities (no ``point-particles'' or
black holes), but this is a serious limitation regarding the
application to compact objects like neutron stars, which can
adequately be approximated by point-masses when studying their
dynamics. Fortunately, the formalism {\it is} applicable to a singular
$T^{\mu\nu}$ involving Dirac measures, at the price of a further
ansatz, that the infinite self-field of point-masses can be
regularized in a certain way. By implementing consistently the
regularization we obtain the multipole moments and the radiation field
of a system of two point-masses at 2.5PN order \cite{BDI95,B96}, as
well as their equations of motion at the same order in the form of
ordinary differential equations \cite{BFP98} (the result agrees with
previous works \cite{BeDD81,DD81a,D82,D83a}); see Section 7.

\section{Multipole decomposition}

In this section we construct the multipole expansion ${\cal
M}(h^{\mu\nu})\equiv h_{\rm ext}^{\mu\nu}$ of the gravitational field
outside an isolated system, supposed to be at once self-gravitating
and slowly-moving. By slowly-moving we mean that the typical current
and stress densities are small with respect to the energy density, in
the sense that
 
\begin{equation}
{\rm max}~\biggl\{ \biggl|{T^{0i}\over T^{00}}\biggr|,
\biggl|{T^{ij}\over T^{00}}\biggr|^{1/2}\biggr\}= O\left({1\over
c}\right)\ ,
\end{equation}
where $1/c$ denotes (slightly abusively) the small post-Newtonian
parameter.  The point about (12) is that the ratio between the size of
the source $d$ and a typical wavelength of the gravitational radiation
is of order $d/\lambda=O(1/c)$. Thus the domain of validity of the
post-Newtonian expansion covers the source: it is given by $r<b$ where
the radius $b$ can be chosen so that $d<b=O(\lambda/c)$.

\subsection{The matching equation}

The construction of the multipole expansion is based on several
technical assumptions, the crucial one being that of the consistency
of the asymptotic matching between the exterior and interior fields of
the isolated system.  In some cases the assumptions can be proved from
the properties of the exterior field $h_{\rm ext}^{\mu\nu}$ as
obtained in \cite{BD86} by means of a post-Minkowskian
algorithm. However, since our assumptions are free of any reference to
the post-Minkowskian expansion, we prefer to state them more
generally, without invoking the existence of such an approximation
(refer to \cite{B98mult} for the full detailed assumptions). In many
cases the assumptions have been explicitly verified at some low
post-Newtonian orders \cite{BD89,DI91a,B95,B96}.

The field $h$ (skipping space-time indices), solution in $I\!\!R^4$ of
the relaxed field equations and the no-incoming radiation condition,
is given as the retarded integral (10). We now {\it assume} that
outside the isolated system, say, in the region $r>{\cal R}$ where
${\cal R}$ is a constant radius strictly larger than $d$, we have
$h={\cal M}(h)$ where ${\cal M}(h)$ denotes the multipole expansion of
$h$, a solution of the {\it vacuum} field equations in $I\!\!R^4$
deprived from the spatial origin $r=0$, and admitting a
spherical-harmonics expansion of a certain structure (see below).
Thus, in $I\!\!R\times I\!\!R^3_*$ where $I\!\!R^3_*\equiv I\!\!R^3-\{
{\bf 0} \}$,

\begin{mathletters}
\begin{eqnarray}
\partial_\nu {\cal M}(h^{\mu\nu})&=&0\ , \\
\Box {\cal M}(h^{\mu\nu}) &=& 
{\cal M}(\Lambda^{\mu\nu})\ .
\end{eqnarray}
\end{mathletters}
The source term ${\cal M}(\Lambda)$ is obtained from inserting ${\cal
M}(h)$ in place of $h$ into (5), i.e. ${\cal
M}(\Lambda)\equiv\Lambda({\cal M}(h))$. [Since the matter tensor has a
compact support, ${\cal M}(T)=0$ so that ${\cal M}(\tau)={c^4\over
16\pi G}{\cal M}(\Lambda)$.]  Of course, inside the source (when
$r\leq d$), the true solution $h$ differs from the vacuum solution
${\cal M}(h)$, the latter becoming in fact singular at the origin
($r=0$). We assume that the spherical-harmonics expansion of ${\cal
M}(h)$ in $I\!\!R\times I\!\!R^3_*$ reads

\begin{equation}
 {\cal M}(h)({\bf x},t) = \sum_{a\leq N} {\hat n}_L r^a (\ln r)^p
 {}_LF_{a,p}(t)+R_N({\bf x},t) \ .
\end{equation}
This expression is valid for any $N\in I\!\!N$. The powers of $r$ are
positive or negative, $a\in Z\!\!\!Z$, and we have $a\leq N$ (the
negative powers of $r$ show that the multipole expansion is singular
at $r=0$). For ease of notation we indicate only the summation over
$a$, but there are two other summations involved: one over the powers
$p\in I\!\!N$ of the logarithms, and one over the order of
multipolarity $l\in I\!\!N$. The summations are considered only in the
sense of formal series, as we do not control the mathematical nature
of the series.  The factor ${\hat n}_L$ is a product of $l$ unit
vectors, $n_L\equiv n^L\equiv n^{i_1}...n^{i_l}$, where $L\equiv
i_1...i_l$ is a multi-index with $l$ indices, on which the symmetric
and trace-free (STF) projection is applied: ${\hat n}_L\equiv{\rm
STF}[n_L]$. The decomposition in terms of STF tensors ${\hat
n}_L(\theta,\varphi)$ is equivalent to the decomposition in usual
spherical harmonics.  The functions ${}_LF_{a,p}(t)$ are smooth
($C^\infty$) functions of time, which become constant when $t\leq
-{\cal T}$ because of our assumption (9).  [Of course, the
${}_LF_{a,p}$'s depend also on $c\,$: ${}_LF_{a,p}(t,c)$.]  Finally
the function $R_N({\bf x},t)$ is defined by continuity throughout
$I\!\!R^4$. Its two essential properties are $R_N\in C^N(I\!\!R^4)$
and $R_N=O(r^N)$ when $r\to 0$ with fixed $t$. In addition $R_N$ is
zero before the time $-{\cal T}$.  Though the function $R_N({\bf
x},t)$ is given ``globally'' (as is the multipole expansion), it
represents a small remainder $O(r^N)$ in the expansion of ${\cal
M}(h)$ when $r\to 0$, which is to be identified with the ``near-zone''
expansion of the field outside the source.  It is convenient to
introduce a special notation for the formal near-zone expansion (valid
to any order $N$):

\begin{equation}
 \overline {{\cal M}(h)}({\bf x},t) = \sum {\hat n}_L r^a (\ln r)^p
 {}_LF_{a,p}(t)\ ,
\end{equation} 
where the summation is to be understood in the sense of formal series.
[Note that (14) and (15) are written for the field variable ${\cal
M}(h)$, but it is easy to check that the same type of structure holds
also for the source term ${\cal M}(\Lambda)$.]

Our justification of the assumed structure (14) is that it has been
{\it proved} to hold for metrics in the class of
Multipolar-post-Minkowskian (MPM) metrics considered in \cite{BD86},
i.e. formal series $h_{\rm ext}=\sum G^nh_n$ which satisfy the vacuum
equations, are stationary in the past, and depend on a {\it finite}
set of independent multipole moments.  More precisely, from the
theorem 4.1 in \cite{BD86}, the general MPM metric $h_{\rm ext}$, that
we identify in this paper with ${\cal M}(h^{\mu\nu})$, is such that
the property (14) holds for the $h_n$'s to any order $n$, with the
only difference that to any finite order $n$ the integers $a,p,l$ vary
into some finite ranges, namely $a_{\rm min}(n)\leq a\leq N$, $0\leq
p\leq n-1$ and $0\leq l\leq l_{\rm max}(n)$, with $a_{\rm min}(n)\to
-\infty$ and $l_{\rm max}(n)\to +\infty$ when $n\to +\infty$.  The
functions ${}_LF_{a,p}$ and the remainder $R_N$ in (14) should
therefore be viewed as post-Minkowskian series $\sum G^n{}_LF_{a,p,n}$
and $\sum G^nR_{N,n}$. What we have done in writing (14) and (15) is
to assume that one can legitimately consider such formal
post-Minkowskian series. Note that because the general MPM metric
represents the most general solution of the field equations outside
the source (Theorem 4.2 in \cite{BD86}), it is quite appropriate to
identify the general multipole expansion ${\cal M}(h)$ with the MPM
metric $h_{\rm ext}$.  Actually we shall justify this assumption in
Section 5 by recovering from ${\cal M}(h)$, step by step in the
post-Minkowskian expansion, the MPM metric $h_{\rm ext}$. Because the
properties are proved in \cite{BD86} for any $n$, and because we
consider the formal post-Minkowskian sum, we see that (14)-(15),
viewed as if it were ``exact'', constitutes a quite natural
assumption. In particular we have assumed in (14)-(15) that the
multipolar series involves an infinite number of independent
multipoles. In summary, we give to the properties (14)-(15) a scope
larger than the one of MPM expansions (maybe they could be proved for
exact solutions), at the price of counting them among our basic
assumptions.

The multipole expansion ${\cal M}(h)$ is a mathematical solution of
the vacuum equations in $I\!\!R\times I\!\!R^3_*$, but whose
``multipole moments'' (the functions ${}_LF_{a,p}$) are not determined
in terms of the source parameters. When the isolated system is slowly
moving in the sense of (12), there exists an overlapping region
between the domains of validity of the post-Newtonian expansion: the
``near-zone'' $r<b$, where $d<b=O(\lambda/c)$, and of the multipole
expansion: the exterior zone $r>{\cal R}$. For this to be true it
suffices to choose ${\cal R}$, which is restricted only to be strictly
larger than $d$, such that $d<{\cal R}<b$. We assume that the field
$h$ given by (10) admits in the near-zone a formal post-Newtonian
expansion, $h={\overline h}$ when $r<b$.  On the other hand, recall
that $h={\cal M}(h)$ when $r > {\cal R}$. Matching the two asymptotic
expansions ${\overline h}$ and ${\cal M}(h)$ in the ``matching''
region ${\cal R}<r<b$ means that the (formal) double series obtained
by considering the multipole expansion of (all the coefficients of)
the post-Newtonian expansion ${\overline h}$ is {\it identical} to the
double series obtained by taking the near-zone expansion of the
multipole expansion. [We use the same overbar notation for the
post-Newtonian and near-zone expansions because the near-zone
expansion ($r/c\to 0$) of the exterior multipolar field is
mathematically equivalent to the expansion when $c\to\infty$ with
fixed multipole moments.] The resulting matching equation reads
 
\begin{equation}
\overline {{\cal M}(h)}={\cal M}(\overline h)\ .
\end{equation} 
This equation should be true term by term, after both sides of the
equation are re-arranged as series corresponding to the same expansion
parameter. Though looking quite reasonable (if the theory makes
sense), the matching equation cannot be justified presently with full
generality; however up to 2PN order it was shown to determine a unique
solution valid everywhere inside and outside the source
\cite{BD89,DI91a,B95}. The matching assumption complements the
framework of MPM approximations \cite{BD86}, by giving physical
``pith'' to the arbitrary multipole moments used in the construction
of MPM metrics (see Section 4).

\subsection{The field in terms of multipole moments}

Let us consider the relaxed vacuum Einstein equation (13b), whose
source term ${\cal M}(\Lambda)$, according to our assumptions, owns
the structure (14) [recall that (14) applies to ${\cal M}(h)$ as well
as ${\cal M}(\Lambda)$]. We obtain a {\it particular} solution of this
equation (in $I\!\!R\times I\!\!R^3_*$) as follows.  First we multiply
each term composing ${\cal M}(\Lambda)$ in (14) by a factor
$(r/r_0)^B$, where $B$ is a complex number and $r_0$ a constant with
the dimension of a length. For each term we can choose the real part
of $B$ large enough so that the term becomes regular when $r\to 0$,
and then we can apply the retarded integral (11).  The resulting
$B$-dependent retarded integral is known to be analytically
continuable for any $B\in I\!\!\!C$ except at integer values including
in general the value of interest $B=0$. Furthermore one can show that
the finite part (in short ${\rm FP}_{B=0}$) of this integral, defined
to be the coefficient of the zeroth power of $B$ in the expansion when
$B\to 0$, is a retarded solution of the corresponding wave equation.
In the case of a regular term in (14) such as the remainder $R_N$,
this solution simply reduces to the retarded integral.  Summing all
these solutions, corresponding to all the separate terms in (14), we
thereby obtain as a particular solution of (13b) the object ${\rm
FP}_{B=0}\, \Box^{-1}_R [ (r/r_0)^B {\cal M}(\Lambda)]$. This is
basically the method employed in \cite{BD86} to solve the vacuum field
equations in the post-Minkowskian approximation.

Now all the problem is to find {\it the} homogeneous solution to be
added to the latter particular solution in order that the multipole
expansion ${\cal M}(h)$ matches with the post-Newtonian expansion
${\overline h}$, solution within the source of the field equation (3)
[or, rather, (10)]. Finding this homogeneous solution means finding
the general consequence of the matching equation (16). The result
\cite{B95,B98mult} is that the multipole expansion $h^{\mu\nu}$
satisfying the Einstein equation (10) together with the matching
equation (16) reads

\begin{equation}    
{\cal M}(h^{\mu\nu}) = {\rm FP}_{B=0}\, \Box^{-1}_R [ (r/r_0)^B {\cal
M}(\Lambda^{\mu\nu})]- {4G\over c^4} \sum^{+\infty}_{l=0} {(-)^l\over
l!} \partial_L \left\{ {1\over r} {\cal H}^{\mu\nu}_L (t-r/c) \right\}
\end{equation} 
where the first term is the previous particular solution, and where
the second term is a retarded solution of the source-free
(homogeneous) wave equation, whose ``multipole moments'' are given
explicitly by ($u\equiv t-r/c$)

\begin{equation}
 {\cal H}^{\mu\nu}_L (u) = {\rm FP}_{B=0}  
\int d^3 {\bf x}~ |{\bf x}/r_0|^B x_L \,
{\overline \tau}^{\mu\nu}({\bf x}, u)\ .
\end{equation} 
Here ${\overline \tau}^{\mu\nu}$ denotes the post-Newtonian expansion
of the stress-energy pseudo-tensor $\tau^{\mu\nu}$ appearing in the
right side of (10).  In (17) and (18) we denote $L=i_1\dots i_l$ and
$\partial_L\equiv \partial_{i_1}\dots\partial_{i_l}$, $x_L\equiv
x_{i_1}\dots x_{i_l}$.

It is important that the multipole moments (18) are found to depend on
the {\it post-Newtonian} expansion ${\overline \tau}^{\mu\nu}$ of the
pseudo-tensor, and not of $\tau^{\mu\nu}$ itself, as this is precisely
where our assumption of matching to the inner post-Newtonian field
comes in. The formula is {\it a priori} valid only in the case of a
slowly-moving source; it is {\it a priori} true only after insertion
of a definite post-Newtonian expansion of the pseudo-tensor, where in
particular all the retardations have been expanded when $c\to\infty$
[the formulas (17)-(18) assume implicitly that one can effectively
construct such a post-Newtonian expansion].

Like in the first term of (17), the moments (18) are endowed with a
finite part operation defined by complex analytic continuation in
$B$. Notice however that the two finite part operations in the first
term of (17) and in (18) act quite differently. In the first term of
(17) the analytic continuation serves at regularizing the singularity
of the multipole expansion at the spatial {\it origin} $r=0$. Since
the pseudo-tensor is smooth inside the source, there is no need in the
moments (18) to regularize the field near the origin; still the finite
part is essential because it applies to the bound of the integral at
{\it infinity} ($|{\bf x}|\to\infty$). Otherwise the integral would be
({\it a priori}) divergent at infinity, because of the presence of the
factor $x_L=O(r^l)$ in the integrand, and the fact that the
pseudo-tensor ${\overline \tau}^{\mu\nu}$ is non-compact
supported. The two finite parts present in the two separate terms of
(17) involve the same arbitrary constant $r_0$, but this constant can
be readily checked to cancel out between the two terms [i.e. the
differentiation of ${\cal M}(h^{\mu\nu})$ with respect to $r_0$ yields
zero].

The formulas (17)-(18) were first obtained (in STF form) up to the 2PN
order in \cite{B95} by performing explicitly the matching. This showed
in particular that the matching equation (16) is correct to 2PN order.
Then the proof valid to any post-Newtonian order, but at the price of
{\it assuming} (16) to all orders, was given in Section 3 of
\cite{B98mult} (see also Appendix A of \cite{B98mult} for an
alternative proof). The crucial step in the proof is to remark that
the finite part of the integral of $\overline{{\cal M}(\Lambda)}$ over
the {\it whole} space $I\!\!R^3$ is identically zero by analytic
continuation:

\begin{equation}
{\rm FP}_{B=0}  
\int_{I\!\!R^3}d^3 {\bf x}~ |{\bf x}/r_0|^B x_L 
\overline{{\cal M}(\Lambda)}({\bf x}, u) = 0\ .
\end{equation}
This follows from the fact that $\overline{{\cal M}(\Lambda)}$ can be
written as a formal series of the type (15). Using (15) it is easy to
reduce the computation of the integral (19) to that of the elementary
radial integral $\int_0^{+\infty} d|{\bf x}||{\bf x}|^{B+2+l+a}$
(since the powers of the logarithm can be obtained by repeatedly
differentiating with respect to $B$). The latter radial integral can
be split into a ``near-zone'' integral, extending from zero to radius
${\cal R}$, and a ``far-zone'' integral, extending from ${\cal R}$ to
infinity (actually any finite non-zero radius fits instead of ${\cal
R}$).  When the real part of $B$ is a large enough positive number,
the value of the near-zone integral is ${\cal R}^{B+3+l+a}/(B+3+l+a)$,
while when the real part of $B$ is a large {\it negative} number, the
far-zone integral reads the opposite, $-{\cal
R}^{B+3+l+a}/(B+3+l+a)$. Both obtained values represent the unique
analytic continuations of the near-zone and far-zone integrals for any
$B\in I\!\!\!C$ except $-3-l-a$.  The complete integral
$\int_0^{+\infty} d|{\bf x}||{\bf x}|^{B+2+l+a}$ is defined as the sum
of the analytic continuations of the near-zone and far-zone integrals,
and is therefore identically zero ($\forall B\in I\!\!\!C$); this
proves (19).

One may ask why the whole integration over $I\!\!R^3$ contributes to
the multipole moment (18) -- a somewhat paradoxical fact because the
integrand is in the form of a post-Newtonian expansion, and is thus
expected to be physically valid (i.e. to give accurate results) only
in the near zone. This fact is possible thanks to the technical
identity (19) which enables us to transform a near-zone integration
into a complete $I\!\!R^3$-integration (refer to \cite{B98mult} for
details).

\subsection{Equivalence with the Will-Wiseman multipole expansion}

Recently a different expression of the multipole decomposition, with
correlatively a different expression of the multipole moments, was
obtained by Will and Wiseman \cite{WWi96}, extending previous work of
Epstein and Wagoner \cite{EW75} and Thorne \cite{Th80}. Basically, the
multipole moments in \cite{WWi96} are defined by an integral extending
over a ball of {\it finite} radius ${\cal R}$ (essentially the same
${\cal R}$ as here), and thus do not require any regularization of the
bound at infinity. By contrast, our multipole moments (18) involve an
integration over the whole $I\!\!R^3$, which is allowed thanks to the
analytic continuation [leading to the identity (19)]. Let us outline
the proof of the equivalence between the Will-Wiseman formalism
\cite{WWi96} and the present one \cite{B95,B98mult}.

Will and Wiseman \cite{WWi96} find, instead of (17)-(18), 

\begin{equation}
{\cal M}(h^{\mu\nu}) = \Box^{-1}_R [{\cal M}(\Lambda^{\mu\nu})]_{|
\cal R} - {4G\over c^4} \sum^{+\infty}_{l=0} {(-)^l\over l!}
\partial_L \left\{ {1\over r} {\cal W}^{\mu\nu}_L (t-r/c) \right\} \ .
\end{equation}
The first term is given by the retarded integral (11) acting on ${\cal
M}(\Lambda)$, but {\it truntated}, as indicated by the subscript
${\cal R}$, to extend only in the ``far zone'': $|{\bf x}'|>{\cal R}$
in the notation (11). Thus, the near-zone part of the retarded
integral, which contains the source, is removed, and there is no
problem with the singularity of the multipole expansion at the
origin. Then, the multipole moments ${\cal W}_L$ are given by an
integral extending over the ``near zone'' only:

\begin{equation}
 {\cal W}^{\mu\nu}_L (u) = \int_{|{\bf x}|<{\cal R}}d^3 {\bf x}~ x_L
 \, {\overline \tau}^{\mu\nu}({\bf x}, u)\ .
\end{equation} 
The integral being compact-supported is well-defined. The multipole
moments ${\cal W}_L$ look technically more simple than ours given by
(18). On the other hand, practically speaking, the analytic
continuation in (18) permits deriving many closed-form formulas to be
used in applications \cite{BDI95,BIJ99}. Of course, one is free to
choose any definition of the multipole moments as far as it is used in
a consistent manner.

We compute the difference between the moments ${\cal H}_L$ and ${\cal
W}_L$. For the comparison we split ${\cal H}_L$ into far-zone and
near-zone integrals corresponding to the radius ${\cal R}$.  Since the
analytic continuation factor in ${\cal H}_L$ deals only with the bound
at infinity, it can be removed from the near-zone integral, which is
then clearly seen to agree with ${\cal W}_L$. So the difference ${\cal
H}_L-{\cal W}_L$ is given by the far-zone integral:

\begin{equation}
 {\cal H}_L(u) - {\cal W}_L(u) ={\rm FP}_{B=0}  
\int_{|{\bf x}|>{\cal R}}d^3 {\bf x}~ |{\bf x}/r_0|^B x_L 
\overline{\tau}({\bf x}, u)\ .
\end{equation}
Next we transform the integrand. Successively we write ${\overline
\tau}={\cal M}({\overline \tau})$ because we are in the far zone;
${\cal M}({\overline \tau})=\overline{{\cal M}(\tau)}$ from the
matching equation (16); and $\overline{{\cal M}(\tau)}= {c^4\over
16\pi G}\overline{{\cal M}(\Lambda)}$ because $T$ has a compact
support. At this stage, the technical identity (19) allows one to
transform the far-zone integration into a near zone integration
(changing simply the overall sign in front of the integral). So,
 
\begin{equation}
 {\cal H}_L (u) - {\cal W}_L (u) 
=-{c^4\over 16\pi G}{\rm FP}_{B=0}  
\int_{|{\bf x}|<{\cal R}}d^3 {\bf x}~ |{\bf x}/r_0|^B x_L 
\overline{{\cal M}(\Lambda)}({\bf x}, u)\ .
\end{equation}
It is straightforward to check that the right side of this equation,
when summed up over all multipolarities $l$, accounts exactly for the
near-zone part that was removed from the retarded integral of ${\cal
M}(\Lambda)$ [first term in (20)], so that the ``complete'' retarded
integral as given by the first term in (17) is exactly
reconstituted. In conclusion the two formalisms \cite{B95,B98mult} and
\cite{WWi96} are equivalent.

\section{Source multipole moments}

Quite naturally our source multipole moments will be closely related
to the ${\cal H}_L$'s obtained in (18). However, before giving a
precise definition, we need to find the equivalent of the multipole
decomposition (17)-(18) in terms of symmetric and trace-free (STF)
tensors, and we must reduce the number of independent tensors by
imposing the harmonic gauge condition (13a). This leads to the
definition of a ``linearized'' metric associated with the multipole
expansion ${\cal M}(h)$, and parametrized by six sets of STF source
multipole moments.

\subsection{Multipole expansion in symmetric-trace-free form}
 
The moments ${\cal H}_L$ given by (18) are non-trace-free because
$x_L$ owns all its traces (i.e. $\delta_{i_li_{l-1}}x_L={\bf
x}^2x_{L-2}$, where $L-2=i_1...i_{l-2}$). Instead of ${\cal H}_L$,
there are certain advantages in using STF multipole moments: indeed
the STF moments are uniquely defined, and they often yield simpler
computations in practice. It is not difficult, using STF techniques,
to obtain the multipole decomposition equivalent to (17)-(18) but
expressed in terms of STF tensors. We find

\begin{equation}
{\cal M}(h^{\mu\nu}) = {\rm FP}_{B=0}\, \Box^{-1}_R [ (r/r_0)^B {\cal
M}(\Lambda^{\mu\nu})] - {4G\over c^4} \sum^{+\infty}_{l=0} {(-)^l\over
l!} \partial_L \left\{ {1\over r} {\cal F}^{\mu\nu}_L (t-r/c) \right\}
\end{equation} 
where the STF multipole moments are given by \cite{B98mult}

\begin{equation}
{\cal F}^{\mu\nu}_L (u) = {\rm FP}_{B=0} \int d^3 {\bf x}~ |{\bf
x}/r_0|^B {\hat x}_L \int^1_{-1} dz~ \delta_l(z) {\overline
\tau^{\mu\nu}} ({\bf x}, u+z|{\bf x}|/c) \ .
\end{equation}
The notation for a STF product of vectors is ${\hat x}_L
\equiv {\rm STF}(x_L)$ (such that ${\hat x}_L$ is symmetric in $L$ and 
$\delta_{i_li_{l-1}}{\hat x}_L=0$; for instance ${\hat x}_{ij}= 
x_i x_j - {1\over 3} \delta_{ij} {\bf x}^2$).
As we see, the STF moments (25) involve an extra integration, over the
variable $z$, with respect to the non-STF ones (18). The weighting
function associated with the $z$-integration reads, for any $l$,

\begin{equation}
\delta_l (z) = {(2l+1)!!\over 2^{l+1} l!} (1-z^2)^l
\ ; \quad\int^1_{-1} dz~\delta_l (z) = 1 \ .
\end{equation} 
In the limit of large $l$ the weighting function tends toward the
Dirac delta measure (hence its name):
$\lim_{l\to\infty}\delta_l=\delta$.  Remark that since (25) is valid
only in the post-Newtonian approximation, the $z$-integration is to be
expressed as a post-Newtonian series. Here is the relevant formula
\cite{BD89}:

\begin{equation}
\int^1_{-1} dz~ \delta_l(z) {\overline \tau}({\bf x}, u+z|{\bf x}|/c)
=\sum_{k=0}^{\infty}{(2l+1)!!\over 2^kk!(2l+2k+1)!!}
\biggl({|{\bf x}|\over c}{\partial\over \partial u}\biggr)^{\!2k}
{\!\overline\tau}({\bf x},u)\ .
\end{equation}
In the limiting case of linearized gravity, one can neglect the first
term in (24), and the pseudo-tensor ${\overline \tau}^{\mu\nu}$ in
(25) can be replaced by the matter stress-energy tensor $T^{\mu\nu}$
(we have ${\overline T}^{\mu\nu}= T^{\mu\nu}$ inside the slowly-moving
source). Since $T^{\mu\nu}$ has a compact support the finite part
prescription can be removed, and we recover the known multipole
decomposition corresponding to a compact-support source (see the
appendix B in \cite{BD89}).

\subsection{Linearized approximation to the exterior field}

Up to now we have solved the {\it relaxed} field equation (10) in the
exterior zone, with result the multipole decomposition (24)-(25).  In
this section we further impose the harmonic gauge condition (13a), and
from this we find a solution of the linearized vacuum equation,
appearing as the first approximation in a post-Minkowskian expansion
of the multipole expansion ${\cal M}(h)$.

Let us give a notation to the first term in (24): 

\begin{equation}
u^{\mu\nu} \equiv {\rm FP}_{B=0}\, \Box^{-1}_R [ (r/r_0)^B
{\cal M}(\Lambda^{\mu\nu})] \ .
\end{equation} 
Applying on (24) the condition $\partial_\nu{\cal M}(h^{\mu\nu})=0$,
we find that the divergence $w^\mu\equiv \partial_\nu u^{\mu\nu}$ is
equal to a retarded solution of the source-free wave equation, given
by

\begin{equation}
w^\mu={4G\over c^4}\partial_\nu\biggl( \sum^{+\infty}_{l=0}
 {(-)^l\over l!} \partial_L \left\{ {1\over r} {\cal
 F}^{\mu\nu}_L (t-r/c) \right\}\biggr)\ .
\end{equation}
Now, associated to any $w^\mu$ of this type, there exists some
$v^{\mu\nu}$ which is like $w^\mu$ a retarded solution of the
source-free wave equation, $\Box (v^{\mu\nu})=0$, and furthermore
whose divergence is the opposite of $w^\mu$, $\partial_\nu
v^{\mu\nu}=-w^\mu$.  We refer to \cite{BD86,B98quad} for the explicit
formulas allowing the ``algorithmic'' construction of $v^{\mu\nu}$
once we know $w^\mu$.  For definiteness, we adopt the formulas (2.12)
in \cite{B98quad}, which represent themselves a slight modification of
the earlier formulas (4.13) in \cite{BD86} (see also the appendix B in
\cite{B98mult}).

With $v^{\mu\nu}$ at our disposal we define what constitutes the
linearized approximation to the exterior metric, say $G h^{\mu\nu}_1$
where we factorize out $G$ in front of the metric in order to
emphasize its linear character:

\begin{equation}
 G h^{\mu\nu}_1 \equiv -{4G\over c^4} \sum^{+\infty}_{l =0} {(-)^l
 \over l !} \partial_L \left\{ {1\over r}{\cal F}^{\mu\nu}_L (t-r/c)
 \right\}- v^{\mu\nu}\ .
\end{equation}
The linearized metric satisfies the linearized vacuum equations in
harmonic gauge: $\Box h^{\mu\nu}_1=0$ since both terms in (30) satisfy
the source-free wave equation, and $\partial_\nu h^{\mu\nu}_1=0$
thanks to (29) and $\partial_\nu v^{\mu\nu}=-w^\mu$. Using the
definition (30) one can re-write the multipole expansion of the
exterior field as

\begin{equation}
 {\cal M} (h^{\mu\nu}) = Gh^{\mu\nu}_1 + u^{\mu\nu}+v^{\mu\nu} \ .
\end{equation}
Quite naturally the $u^{\mu\nu}$ and $v^{\mu\nu}$ will represent the
{\it non-linear} corrections to be added to the ``linearized'' metric
$Gh^{\mu\nu}_1$ in order to reconstruct the complete exterior metric
(see Section 4).

Since $h^{\mu\nu}_1$ satisfies $\Box h^{\mu\nu}_1= 0= \partial_\nu
h^{\mu\nu}_1$, there is a unique way to decompose it into the sum of a
``canonical'' metric introduced by Thorne \cite{Th80} (see also
\cite{BD86}) plus a linearized gauge transformation,
 
\begin{equation}
 h^{\mu\nu}_1 = h^{\mu\nu}_{\rm can1} + \partial^\mu\varphi^\nu_1 +
 \partial^\nu\varphi^\mu_1 - \eta^{\mu\nu}
 \partial_\lambda\varphi^\lambda_1\ .
\end{equation}
The canonical linearized metric is defined by
 
\begin{mathletters}
\begin{eqnarray}
 h^{00}_{\rm can1} &=& -{4\over c^2}\sum_{l\geq 0} {(-)^l\over l !}
 \partial_L \left( {1\over r} I_L (u)\right)\ ,\\ h^{0i}_{\rm can1}
 &=& {4\over c^3}\sum_{l\geq 1} {(-)^l\over l !} \left\{
 \partial_{L-1} \left( {1\over r} I_{iL-1}^{(1)} (u)\right)
 \right.\nonumber \\ && \qquad\qquad\qquad +\left.  {l\over l+1}
 \varepsilon_{iab} \partial_{aL-1} \left( {1\over r} J_{bL-1}
 (u)\right)\right\} \ , \\ h^{ij}_{\rm can1} &=&-{4\over
 c^4}\sum_{l\geq 2}\! {(-)^l\over l !}\! \left\{ \partial_{L-2} \left(
 {1\over r} I_{ijL-2}^{(2)} (u)\right)\right. \nonumber\\ &&
 \qquad\qquad\qquad \left.+ {2l\over l+1} \partial_{aL-2} \left(
 {1\over r} \varepsilon_{ab(i} J_{j)bL-2}^{(1)} (u)\right)\right\} \ ,
\end{eqnarray}
\end{mathletters}
where the $I_L$'s and $J_L$'s are two sets of functions of the
retarded time $u=t-r/c$ [the subscript $(n)$ indicates $n$ time
derivatives], and which are STF with respect to all their indices
$L=i_1\dots i_l$ (the symmetrization is denoted with parenthesis). As
for the gauge vector $\varphi^\mu_1$, it satisfies
$\Box\varphi^\mu_1=0$ and depends in a way similar to (33) on four
other sets of STF functions of $u$, denoted $W_L$, $X_L$, $Y_L$ and
$Z_L$ (one type of function for each component of the vector). See
\cite{B98mult} for the expression of
$\varphi^\mu_1=\varphi^\mu_1[W_L,X_L,Y_L,Z_L]$.

\subsection{Derivation of the source multipole moments}

The two sets of multipole moments $I_L$ and $J_L$ parametrizing the
metric (33) constitute our definitions for respectively the mass-type
and current-type multipole moments of the source. Actually, there are
also the moments $W_L$, $X_L$, $Y_L$, $Z_L$, and we refer collectively
to $\{I_L,J_L,W_L,X_L,Y_L,Z_L\}$ as the set of six {\it source}
multipole moments.

With (32) it is easily seen (because $\Box\varphi^\mu_1=0$) that the
gauge condition $\partial_\nu h^{\mu\nu}_1=0$ imposes no condition on
the source moments except the conservation laws appropriate to the
gravitational monopole $I$ (having $l=0$) and dipoles $I_i$, $J_i$
($l=1$): namely,

\begin{equation}
I^{(1)}=0~;\qquad I_i^{(2)}=0~;\qquad J_i^{(1)}=0\ .
\end{equation}
The mass monopole $I$ and current dipole $J_i$ are thus constant, and
agree respectively with the ADM mass and total angular momentum of the
isolated system (later we shall denote the ADM mass by $M\equiv
I$). According to (34) the mass dipole $I_i$ is a linear function of
time, but since we assumed that the metric is stationary in the past,
$I_i$ is in fact also constant, and equal to the (ADM) center of mass
position.

The expressions of $I_L$ and $J_L$ (as well as of the other moments
$W_L,X_L,Y_L,$ $Z_L$) come directly from (30) with (32)-(33) and the
result of the matching, which is personified by the formula (25). To
simplify the notation we define

\begin{mathletters}
\begin{eqnarray}
  {\Sigma} &\equiv& {\overline\tau^{00} +\overline\tau^{ii}\over c^2}\
  ,\\ {\Sigma}_i &\equiv& {\overline\tau^{0i}\over c}\ ,\\
  {\Sigma}_{ij} &\equiv& \overline{\tau}^{ij}\ ,
\end{eqnarray}
\end{mathletters}
(where $\overline{\tau}^{ii} \equiv\delta_{ij}\overline\tau^{ij}$). 
The result is \cite{B98mult} 
\begin{mathletters}
\begin{eqnarray}
 I_L(u)&=& \hbox{FP}_{B=0} \int\! d^3{\bf x}~|{\bf x}/r_0|^B
 \int^1_{-1} dz\biggl\{ \delta_l\hat x_L \Sigma -{4(2l+1)\over
  c^2(l+1)(2l+3)} \delta_{l+1} \hat x_{iL} \partial_t{\Sigma}_i
  \nonumber\\
 &&\qquad +{2(2l+1)\over c^4(l+1)(l+2)(2l+5)} \delta_{l+2} \hat x_{ijL}
  \partial_t^2{\Sigma}_{ij} \biggr\} ({\bf x},u+z |{\bf x}|/c)\ ,
  \\ \nonumber \\
J_L(u)&=& \varepsilon_{ab<i_l} \hbox{FP}_{B=0} \int
   d^3{\bf x}~|{\bf x}/r_0|^B \int^1_{-1} dz\biggl\{ \delta_l\hat
  x_{L-1>a}  \Sigma_b   \nonumber\\ 
&&\qquad -{2l+1\over c^2(l+2)(2l+3)} \delta_{l+1} \hat x_{L-1>ac}    
  \partial_t{\Sigma}_{bc}
  \biggr\} ({\bf x},u+z |{\bf x}|/c)\ , 
\end{eqnarray}
\end{mathletters}
($<>$ refers to the STF projection). In a sense these expressions are
{\it exact}, since they are formally valid up to any post-Newtonian
order.  [See (68)-(69) below for explicit formulas at 2PN.]

By replacing ${\overline\tau}^{\mu\nu}$ in (36) by the compact-support
matter tensor $T^{\mu\nu}$ we recover the expressions of the multipole
moments worked out in linearized gravity by Damour and Iyer
\cite{DI91b} (see also \cite{CMM77}). On the other hand the formulas
(36) contain the results obtained by explicit implementation (``order
by order'') of the matching up to the 2PN order \cite{B95}.

\section{Post-Minkowskian approximation}

In linearized gravity, the source multipole moments represent also the
moments which are ``measured'' at infinity, using an array of
detectors surrounding the source. However, in the non-linear theory,
the gravitational source $\Lambda^{\mu\nu}$ cannot be neglected and
the first term in (24) plays a crucial role, notably it implies that
the measured multipole moments at infinity differ from the source
moments.  Thus, we must now supplement the formulas of the source
multipole moments (36) by the study of the ``non-linear'' term
$u^{\mu\nu}\equiv {\rm FP}_{B=0} \Box^{-1}_R [(r/r_0)^B {\cal
M}(\Lambda^{\mu\nu})]$ in (24). For this purpose we develop following
\cite{BD86} a post-Minkowskian approximation for the exterior vacuum
metric.

\subsection{Multipolar-post-Minkowskian iteration of the exterior field}

The work started already with the formulas (31)-(33), where we
expressed the exterior multipolar metric $h_{\rm ext}^{\mu\nu}\equiv
{\cal M}(h^{\mu\nu})$ as the sum of the ``linearized'' metric
$Gh_1^{\mu\nu}$ and the ``non-linear'' corrections $u^{\mu\nu}$, given
by (28), and $v^{\mu\nu}$, algorithmically constructed from $w^\mu =
\partial_\nu u^{\mu\nu}$ [see (29)]. The linearized metric is a
functional of the source multipole moments: $h_1=
h_1[I,J,W,X,Y,Z]$. We regard $G$ as the book-keeping parameter for the
post-Minkowskian series, and consider that $Gh_1$ is purely of first
order in $G$, and thus that $h_1$ itself is purely of zeroth order. Of
course we know from the previous section that this is untrue, because
the source multipole moments depend on $G$; supposing $h_1=O(G^0)$ is
simply a convention allowing the systematic implementation of the
post-Minkowskian iteration.

Here we check that the non-linear corrections $u^{\mu\nu}$ and
$v^{\mu\nu}$ in (31) generate the whole post-Minkowskian algorithm of
\cite{BD86}. The detail demanding attention is how the
post-Minkowskian expansions of $u^{\mu\nu}$ and $v^{\mu\nu}$ are
related to a spliting of the gravitational source $\Lambda^{\mu\nu}$
into successive non-linear terms. Let us pose, with obvious notation,

\begin{equation}
\Lambda^{\mu\nu}=N^{\mu\nu}[h,h]+M^{\mu\nu}[h,h,h]+O(h^4)\ ,
\end{equation}
where, from the exact formula (5), the quadratic-order piece reads
(all indices being lowered with the Minkowski metric, and $h$ denoting
$\eta^{\rho\sigma}h_{\rho\sigma}$):

\begin{eqnarray}
 N^{\mu\nu}[h,h]=&-& h^{\rho\sigma} \partial^2_{\rho\sigma} h^{\mu\nu}
 + {1\over 2} \partial^\mu h_{\rho\sigma} \partial^\nu h^{\rho\sigma}
 - {1\over 4} \partial^\mu h \partial^\nu h \nonumber\\
 &-&\partial^{\mu} h_{\rho\sigma} \partial^\rho h^{\nu\sigma}
 -\partial^{\nu} h_{\rho\sigma} \partial^\rho h^{\mu\sigma}
 +\partial_\sigma h^{\mu\rho} (\partial^\sigma h^\nu_\rho +
 \partial_\rho h^{\nu\sigma}) \nonumber\\ &+& \eta^{\mu\nu} \biggl[
 -{1\over 4}\partial_\lambda h_{\rho\sigma} \partial^\lambda
 h^{\rho\sigma} +{1\over 8}\partial_\rho h \partial^\rho h +{1\over
 2}\partial_\rho h_{\sigma\lambda} \partial^\sigma
 h^{\rho\lambda}\biggr]\ ,
\end{eqnarray}
and where the cubic-order piece $M[h,h,h]$ and all higher-order terms
can be obtained in a straightforward way.

First, reasoning {\it ad absurdio}, we prove (see \cite{B98mult} for
details) that both $u$ and $v$ indeed represent non-linear corrections
to the linearized metric since they start at order $G^2$:
$u=G^2u_2+O(G^3)$ and $v=G^2v_2+O(G^3)$. Next we obtain explicitly
$u_2$ by substituting the linearized metric $h_1$ into (38) and
applying the finite part of the retarded integral, i.e.

\begin{equation}
u^{\mu\nu}_2={\rm FP}_{B=0}\Box_R^{-1}\biggl\{(r/r_0)^B
N^{\mu\nu}[h_1,h_1]\biggr\}\ .
\end{equation}
In this way we have a particular solution of the wave equation in
$I\!\!R\times I\!\!R^3_*$, $\Box u_2= N[h_1,h_1]$. From $u_2$ one
deduces $v_2$ by the same ``algorithmic'' equations as used when
deducing $v$ from $u$ [see after (29)]. Then $\Box v_2=0$ and the sum
$u_2+v_2$ is divergenceless, so we can solve the quadratic-order
vacuum equations in harmonic coordinates by posing

\begin{equation}
h^{\mu\nu}_2=u^{\mu\nu}_2+v^{\mu\nu}_2\ .
\end{equation}
With this definition it is clear that the multipole expansion (31)
reads to quadratic order:

\begin{equation}
 {\cal M} (h^{\mu\nu}) = Gh^{\mu\nu}_1 + G^2h^{\mu\nu}_2+O(G^3)\ .  
\end{equation}
Continuing in this fashion to the next order we find successively

\begin{mathletters}
\begin{eqnarray}
&& u^{\mu\nu}_3={\rm FP}_{B=0}\Box_R^{-1}\biggl\{\!\!(r/r_0)^B\!
\biggl(\!M^{\mu\nu}[h_1,h_1,h_1]+N^{\mu\nu}[h_1,h_2]+N^{\mu\nu}[h_2,h_1]
\biggr)\!\biggr\} ;\nonumber \\ && \\
&& h^{\mu\nu}_3=u^{\mu\nu}_3+v^{\mu\nu}_3 ;\\ &&\nonumber \\
&& {\cal M} (h^{\mu\nu})=Gh^{\mu\nu}_1 + G^2h^{\mu\nu}_2+G^3h^{\mu\nu}_3+O(G^4)  
\ .
\end{eqnarray}
\end{mathletters}
This process continues {\it ad infinitum}.  The latter
post-Minkowskian algorithm is exactly the one proposed in \cite{BD86}
(see also Section 2 of \cite{B98quad}). That is, starting from
$h_1[I,J,W,X,Y,Z]$ given by (32)-(33), one generates the infinite
post-Minkowskian (MPM) series of \cite{BD86}, solving the vacuum
(harmonic-coordinate) Einstein equations in $I\!\!R\times I\!\!R^3_*$,
and this formal series happens to be equal, term by term in $G$, to
the {\it general} multipole decomposition of $h^{\mu\nu}$ given by
(24). For any $n$, we have $h^{\mu\nu}_n=u^{\mu\nu}_n+v^{\mu\nu}_n$,
and

\begin{equation}
{\cal M}(h^{\mu\nu})=\sum_{n=1}^{+\infty}G^nh^{\mu\nu}_n \ .
\end{equation}
This result is perfectly consistent with the fact that the MPM
algorithm generates the most {\it general} solution of the field
equations in $I\!\!R\times I\!\!R^3_*$. Furthermore, the latter
post-Minkowskian approximation is known \cite{DSch90} to be reliable
(existence of a one-parameter family of exact solutions whose Taylor
expansion when $G\to 0$ reproduces the approximation) --~an
interesting result which indicates that the multipole decomposition
${\cal M}(h)$ given by (24)-(25) might be proved within a context of
exact solutions.

Recall that the source multipole moments $I_L$, $J_L$, $W_L$, $X_L$,
$Y_L$, $Z_L$ entering the linearized metric $h_1$ at the basis of the
post-Minkowskian algorithm are given by formulas like (36). Thus, in
the present formalism, the source moments, including formally all
post-Newtonian corrections [and all possible powers of $G$] as
contained in (36), serve as ``seeds'' for the post-Minkowskian
iteration of the exterior field, which as it stands leads to all
possible non-linear interactions between the moments. As we can
imagine, rapidly the formalism becomes extremely complicated when
going to higher and higher post-Minkowskian and/or post-Newtonian
approximations. Most likely the complexity is not due to the formalism
but reflects the complexity of the field equations.  It is probably
impossible to find a different formalism in which things would be much
simpler (except if one restricts to a particular type of source).

\subsection{The ``canonical'' multipole moments}

The previous post-Minkowskian algorithm started with $h_1$, a
functional of {\it six} types of source multipole moments, $I_L$ and
$J_L$ entering the ``canonical'' linearized metric $h_{\rm can1}$
given by (33), and $W_L$, $X_L$, $Y_L$, $Z_L$ parametrizing the gauge
vector $\varphi_1$ in (32). All these moments deserve their name of
source moments, but clearly the moments $W_L$, $X_L$, $Y_L$ and $Z_L$
do not play a physical role at the level of the linearized
approximation, as they simply parametrize a linear gauge
transformation. But because the theory is covariant with respect to
(non-linear) diffeomorphisms and not merely to linear gauge
transformations, these moments do contribute to physical quantities at
the non-linear level.

In practice, the presence of the moments $W_L$, $X_L$, $Y_L$, $Z_L$
complicates the post-Minkowskian iteration. Fortunately one can take
advantage of the fact (proved in \cite{BD86}) that it is always
possible to parametrize the vacuum metric by means of two and only two
types of multipole moments $M_L$ and $S_L$ (different from $I_L$ and
$J_L$). The metric is then obtained by the same post-Minkowskian
algorithm as in (39)-(43), but starting with the ``canonical''
linearized metric $h_{\rm can1}[M,S]$ instead of
$h_1[I,J,W,X,Y,Z]$. The resulting non-linear metric $h_{\rm can}$ is
isometric to our exterior metric $h_{\rm ext}\equiv {\cal M}(h)$,
provided that the moments $M_L$ and $S_L$ are given in terms of the
source moments $I_L,J_L,\dots,Z_L$ by some specific relations

\begin{mathletters}
\begin{eqnarray}
 M_L &=& M_L [ I, J, W, X, Y, Z]\ ,\\ S_L &=& 
S_L [ I, J, W, X, Y, Z]\ .
\end{eqnarray}
\end{mathletters}
The two coordinate systems in which $h_{\rm can}$ and $h_{\rm ext}$
are defined satisfy the harmonic gauge condition in the exterior zone,
but (probably) only the one associated with $h_{\rm ext}$ meshes with
the harmonic coordinates in the interior zone. With the notation (32)
the coordinate change reads $\delta x^\mu=G\varphi_1^\mu$+ non-linear
corrections. We shall refer to the moments $M_L$ and $S_L$ as the
mass-type and current-type {\it canonical} multipole moments.  Of
course, since at the linearized approximation the only ``physical''
moments are $I_L$ and $J_L$, we have

\begin{mathletters}
\begin{eqnarray}
 M_L &=& I_L + O(G)\ ,\\ S_L &=& J_L + O(G)\ ,
\end{eqnarray}
\end{mathletters}
where $O(G)$ denotes the post-Minkowskian corrections. Furthermore, it
can be shown \cite{B96} that in terms of a post-Newtonian expansion
the difference between both types of moments is very small: 2.5PN
order, i.e.
 
\begin{equation}
M_L=I_L+O\left({1\over c^5}\right)
\end{equation}
[note that $M=M_{\rm ADM}=I$]. Thus, from (46), the canonical moments
are only ``slightly'' different from the source moments. Their
usefulness is merely practical --~in general they are used in place of
the source moments to simplify a computation.

\subsection{Retarded integral of a multipolar extended source}

The previous post-Minkowskian algorithm has only theoretical interest
unless we supply it with some {\it explicit} formulas for the
computation of the coefficients $h_n$. Happily for us pragmatists,
such formulas exist, and can be found in a rather elegant way thanks
to the process of analytic continuation. Basically we need the
retarded integral of an extended (non-compact-support) source with a
definite multipolarity $l$. Here we present three exemplifying
formulas; see the appendices A in \cite{B98quad} and \cite{B98tail}
for more discussion.

Very often we meet a wave equation whose source term is of the type
${\hat n_L}F(t-r/c)/r^k$, where ${\hat n_L}$ has multipolarity $l$ and
$F$ denotes a certain product of multipole moments. [Clearly, the
near-zone expansion of such a term is of the form (15).] When the
power $k$ is such that $3\leq k\leq l+2$ (this excludes the scalar
case $l=0$), we obtain the solution of the wave equation as
\cite{BD86,BD88}

\begin{eqnarray}
&&\hbox{FP}_{B=0} \Box^{-1}_R \biggl[(r/r_0)^B {\hat n_L\over r^k}
F(t-r/c) \biggr]= -{(k-3)!(l+2-k)!\over (l+k-2)!}{\hat n_L}
\nonumber\\ &&\qquad\qquad\qquad \qquad \quad \times
\sum^{k-3}_{j=0}2^{k-3-j}{(l+j)!\over j!(l-j)!}
{F^{(k-3-j)}(t-r/c)\over c^{k-3-j}~r^{j+1}}\ .
\end{eqnarray} 
As we see the (finite part of the) retarded integral depends in this
case on the values of the extended source at the {\it same} retarded
time $t-r/c$ (for simplicity we use the same notation for the source
and field points). But it is well known (see e.g. \cite{BoR66,HR69})
that this feature is exceptional; in most cases the retarded integral
depends on the whole integrated past of the source. A chief example of
such a ``hereditary'' character is the case with $k=2$ in the previous
example, for which we find \cite{BD88,BD92}
 
\begin{equation}
\Box^{-1}_R \biggl[ {\hat n_L \over r^2} F(t-r/c) \biggr] 
= - {\hat n_L\over r} \int^{ct-r}_{-\infty} ds F(s/c)
Q_l\biggl({ct-s\over r}\biggr)\
\end{equation}
where $Q_l$ denotes the Legendre function of the second kind, related
to the usual Legendre polynomial $P_l$ by the formula

\begin{equation}
 Q_l(x) = {1\over 2} P_l (x) {\rm ln} \left({x+1 \over x-1} \right)-
 \sum^l_{ j=1} {1 \over j} P_{l -j}(x) P_{j-1}(x)\ .
\end{equation}
Since the retarded integral (48) is in fact convergent when $r\to 0$,
we have removed the factor $(r/r_0)^B$ and finite part
prescription. When the source term itself is given by a ``hereditary''
expression such as the right side of (48), we get a more complicated
but still manageable formula, for instance \cite{B98tail}

\begin{equation}
\Box^{-1}_R \biggl[
{\hat{n}_L\over r^2} \int^{ct-r}_{-\infty} \!\!\!ds F(s/c)
Q_p\biggl({ct-s\over r}\biggr) \biggr]= {c\hat n_L\over r}
\int^{ct-r}_{-\infty}\!\!\! ds F^{(-1)}(s/c) R_{lp}\biggl({ct-s\over
r}\biggr)
\end{equation}
where $F^{(-1)}$ denotes that anti-derivative of $F$ which is zero in
the past [from (9) we have restricted $F$ to be zero in the past], and
where

\begin{equation}
R_{lp}(x)=Q_l (x) \int^x_1 dy\, Q_p (y) {dP_l\over dy} (y) + P_l (x)
\int^{+\infty}_x dy\, Q_p (y) {dQ_l\over dy} (y) \ .
\end{equation}
Like in (48) we do not need a finite part operation. The function
$R_{lp}$ is well-defined thanks to the behaviour of the Legendre
function at infinity: $Q_l(x)\sim 1/x^{l+1}$ when $x\to\infty$.

The formulas (48)-(51) are needed to investigate the so-called tails
of gravitational waves appearing at quadratic non-linear order, and
even the tails generated by the tails themselves (``tails of tails'')
which arise at cubic order \cite{BD92,B98tail}. (These formulas do not
show a dependence on the constant $r_0$, but other formulas do.)

\section{Radiative multipole moments}

In Section 2 we introduced the {\it definition} of a set of multipole
moments $\{I_L,J_L,W_L,X_L,Y_L,Z_L\}$ for the isolated source, and in
Section 3 we showed that the exterior field, and in particular the
asymptotic field therein, is actually a complicated non-linear
functional of the latter moments. Therefore, to define some source
multipole moments is not sufficient by itself; this must be completed
by a study of the relation between the adopted definition and some
convenient far-field observables. The same is true of other
definitions of source moments in different formalisms, such as in the
Dixon local description of extended bodies
\cite{Dixon79,EhlRu77,Schatt79}, which should be completed by a
connection to the far-zone gravitational field, for instance along the
line proposed by \cite{SchattSt81,StSchatt81} in the case of the Dixon
moments. In the present formalism, the connection rests on the
relation between the so-called {\it radiative} multipole moments,
denoted $U_L$ and $V_L$, and the source moments $I_L$, $J_L$,$\dots$,
$Z_L$ [in fact, for simplicity's sake, we prefer using the two moments
$M_L$ and $S_L$ instead of the more basic six source moments].

\subsection{Definition and general structure}

The radiative moments $U_L$ (mass-type) and $V_L$ (current-type) are
the coefficients of the multipolar decomposition of the leading $1/R$
part of the transverse-tracefree (TT) projection of the radiation
field in radiative coordinates $(T,{\bf X})$ (with $R=|{\bf X}|$ the
radial distance to the source). Radiative coordinates are such that
the metric coefficients admit an expansion when $R\to\infty$ in powers
of $1/R$ (no logarithms of $R$). In radiative coordinates the retarded
time $T-R/c$ is light-like, or becomes asymptotically light-like when
$R\to\infty$. By {\it definition},

\begin{eqnarray}
 h^{TT}_{ij} ({\bf X},T)&=&{4G\over c^2R} {\cal P}_{ijab} ({\bf N})
 \sum_{l \geq 2} {1\over c^l l !} \biggl\{ N_{L-2} U_{abL-2}\nonumber
 \\ &&\qquad\quad - {2l \over c(l+1)} N_{cL-2} \varepsilon_{cd(a}
 V_{b)dL-2}\biggr\}+ O\left( {1\over R^2}\right)\ ,
\end{eqnarray}
where $N_i=X^i/R$, $N_{L-2}=N_{i_1}\dots N_{i_{l-2}}$,
$N_{cL-2}=N_cN_{L-2}$, and the TT {\it algebraic} projector reads
${\cal P}_{ijab}= (\delta_{ia} -N_iN_a)(\delta_{jb} -N_jN_b) -{1\over
2} ( \delta_{ij} -N_iN_j) (\delta_{ab}-N_aN_b)$.  The radiative
moments $U_L$ and $V_L$ depend on $T-R/c\,$; from (52) they are
defined $\forall ~l\geq 2$. The radiative-coordinate retarded time
differs from the corresponding harmonic-coordinate time by the
well-known logarithmic deviation of light cones,

\begin{equation}
T-{R\over c}=t-{r\over c}-{2GM\over c^3}\ln\left({r\over r_0}\right)
+O(G^2) \ ,
\end{equation}
where we have introduced in the logarithm the same constant $r_0$ as
in (39) (this corresponds simply to a choice of the origin of time in
the far zone).

Now from the post-Minkowskian algorithm of Section 3, it is clear that
the radiative moments $U_L$ and $V_L$ can be obtained to any
post-Minkowskian order in principle, in the form of a non-linear
series in the source or equivalently the canonical multipole moments
$M_L$ and $S_L$.  The practical detail (worked out in \cite{B87}) is
to determine the transformation between harmonic and radiative
coordinates, generalizing (53) to any post-Minkowskian order. The
structure of e.g. the mass-type radiative moment is

\begin{equation}
 U_L =M_L^{(l)}+ \sum_{n=2}^{+\infty} {G^{n-1}\over
  c^{3(n-1)+2k}} X_{nL}\ .
\end{equation}
The first term comes from the fact that the radiative moment reduces
at the linearized approximation to the ($l$th time derivative of the)
source or canonical moment. The second term represents the series of
non-linear corrections, each of them is given by a certain $X_{nL}$
which is a $n$-linear functional of derivatives of multipole moments
$M_L$ or $S_L$. Furthermore we know from e.g. (48) and (50) that each
new non-linear iteration (which always involves a retarded integral)
brings {\it a priori} a new ``hereditary'' integration with respect to
the previous approximation. So we expect that $X_{nL}$ is of the form
($U\equiv T-R/c$)

\begin{equation}
  X_{nL}(U)=\sum \int_{-\infty}^U\!\!du_1\dots\int_{-\infty}^U\!\!du_n
  {\cal Z}_n(U,u_1,\dots,u_n) M^{(a_1)}_{L_1}(u_1)\dots
  S^{(a_n)}_{L_n}(u_n)
\end{equation}
where ${\cal Z}_n$ denotes a certain kernel depending on time
variables $U,u_1,\dots,u_n$, and where the sum refers to all
possibilities of coupling together the $n$ moments. [See (56) below
for examples of kernels ${\cal Z}_2$ and ${\cal Z}_3$.] A useful
information is obtained from imposing that ${\cal Z}_n$ be
dimensionless; this yields the powers of $G$ and $1/c$ in front of
each non-linear term in (54), where $k$ is the number of contractions
among the indices present on the $n$ moments (the current moments
carrying their associated Levi-Civita symbol).

As an example of application of (54) let us suppose that one is
interested in the 3PN or $1/c^6$ approximation. From (54) we have
$3(n-1)+2k=6$, and we deduce that the only possibility is $n=3$ (cubic
non-linearity) and $k=0$ (no contractions between the moments).  From
this we infer immediately that the only possible multipole interaction
at that order is between two mass monopoles and a multipole,
i.e. $M\times M\times M_L$. This corresponds to the ``tails of tails''
computed explicitly in (56) below.
 
\subsection{The radiative quadrupole moment to 3PN order}

To implement the formula (54) a tedious computation is to be done,
following in details the post-Minkowskian algorithm of Section 4
augmented by explicit formulas such as (47)-(51), and changing the
coordinates from harmonic to radiative according to the prescription
in \cite{B87}.  Here we present the result of the computation of the
mass-type radiative quadrupole ($l=2$) up to the 3PN order:

\begin{eqnarray}
&&U_{ij}(U) = M^{(2)}_{ij}(U) + 2 {GM\over c^3} \int^{+\infty}_0\!\!\! d
\tau M^{(4)}_{ij} (U-\tau) \left[ \ln \left( {c\tau\over 2r_0} \right)
+ {11\over 12} \right] \nonumber\\ &&\quad + {G\over c^5} \biggl\{ -
{2\over 7} \int^{+\infty}_0 \!\!\! d \tau
\left[M^{(3)}_{a<i} M^{(3)}_{j>a}\right](U-\tau) - {2\over 7} M^{(3)}_{a<i}
M^{(2)}_{j>a} \nonumber\\ &&\quad - {5\over 7} M^{(4)}_{a<i}
M^{(1)}_{j>a} + {1\over 7} M^{(5)}_{a<i} M_{j>a} + {1 \over 3}
\varepsilon_{ab<i} M^{(4)}_{j>a} S_b \biggr\} \nonumber\\ &&\quad + 2
\left( {GM\over c^3} \right)^2\!\!\! \int^{+\infty}_0\!\! d
\tau M^{(5)}_{ij}(U-\tau)
\left[ \ln^2 \left( {c\tau\over 2r_0} \right) + {57\over 70}
\ln \left( {c\tau\over 2r_0} \right) + {124627\over 44100} \right] 
\nonumber\\
&&\quad + O \left( {1\over c^7} \right)\ .
\end{eqnarray}
Recall that in this formula the moment $M_{ij}$ is the canonical
moment which agrees with the source moment $I_{ij}$ up to a 2.5PN term
[see (46)], and that the source moment $I_{ij}$ itself is given in
terms of the pseudo-tensor of the source by (36a). See also the
formulas (68)-(69) below for a more explicit expression of the source
moment at the 2PN order [of course, to be consistent, one should use
(56) conjointly with 3PN expressions of the source moments].

The ``Newtonian'' term in (56) corresponds to the quadrupole
formalism.  Next, there is a quadratic non-linear correction with
multipole interaction $M\times M_{ij}$ representing the dominant
effect of tails (scattering of linear waves off the space-time
curvature generated by the mass $M$). This correction, computed in
\cite{BD92}, is of order $1/c^3$ or 1.5PN and has the form of a
hereditary integral with logarithmic kernel. The constant $11/12$
depends on the coordinate system chosen to cover the source, here the
harmonic coordinates; for instance the constant would be $17/12$ in
Schwarzschild-like coordinates \cite{P93,BS93}.  The next correction,
of order $1/c^5$ or 2.5PN, is constituted by quadratic interactions
between two mass quadrupoles, and between a mass quadrupole and a
constant current dipole \cite{B98quad}.  This term contains a
hereditary integral, of a type different from the tail integral, which
is due to the gravitational radiation generated by the stress-energy
distribution of linear waves \cite{P83,B90,WiW91,BD92}. Sometimes this
integral is referred to as the non-linear memory integral because it
corresponds to the contribution of gravitons in the so-called linear
memory effect \cite{Th92}. The non-linear memory integral can easily
be found by using the effective stress-energy tensor of gravitational
waves in place of the right side of (3); it follows also from rigorous
studies of the field at future null infinity
\cite{Chr91,Fr92}. Finally, at 3PN order in (56) appears the dominant
cubic non-linear correction, corresponding to the interaction $M\times
M\times M_{ij}$ and associated with the tails of tails of
gravitational waves \cite{B98tail}.

\subsection{Tail contributions in the total energy flux}

Observable quantities at infinity are expressible in terms of the
radiative mass and current multipole moments. For instance the total
gravitational-wave power emitted in all spatial directions (total
gravitational flux or ``luminosity'' ${\cal L}$) is given by the
positive-definite multipolar series

\begin{eqnarray}
 {\cal L} &=& \sum^{+\infty}_{l=2} {G\over c^{2l+1}} \biggl\{
 {(l+1)(l+2)\over l(l-1)l!(2l+1)!!} U^{(1)}_L U^{(1)}_L \nonumber \\
 &&\qquad\qquad\qquad + {4l (l+2)\over c^2(l-1)(l+1)!(2l+1)!!}
 V^{(1)}_L V^{(1)}_L \biggr\}\ .
\end{eqnarray}
In the case of inspiralling compact binaries (a most prominent source
of gravitational waves) the rate of inspiral is fixed by the flux
${\cal L}$, which is therefore a crucial quantity to
predict. Excitingly enough, we know that ${\cal L}$ should be
predicted to 3PN order for detection and analysis of inspiralling
binaries in future experiments \cite{3mn,CFPS93}.

To 3PN order we can use the relation (56) giving the 3PN radiative
quadrupole moment. Here we concentrate our attention on tails and
tails of tails.  The dominant tail contribution at 1.5PN order yields
correspondingly a contribution in the total flux (with $U=T-R/c$):

\begin{equation}
{\cal L}_{\rm tail}={4G^2M\over 5c^8}I^{(3)}_{ij}(U)
\int^{+\infty}_0\! d\tau I^{(5)}_{ij} (U-\tau) \left[\ln
\left({c\tau \over 2r_0}\right) +{11\over 12}\right]\ .
\end{equation}
Since we are interested in the dominant tail we have replaced using
(46) the canonical mass quadrupole by the source quadrupole. Similarly
there are some tail contributions due to the mass octupole, current
quadrupole and all higher-order multipoles, but these are
correlatively of higher post-Newtonian order [see the factors $1/c$ in
(57)]. It has been shown \cite{BD88} that the work done by the
dominant ``hereditary'' contribution in the radiation reaction force
within the source --~which arises at 4PN order in the equations of
motion~-- agrees exactly with (58).

Next, because ${\cal L}$ is made of squares of (derivatives of)
radiative moments, it contains a term with the square of the tail
integral at 1.5PN. This term arises at the 3PN relative order and 
reads

\begin{equation}
{\cal L}_{(\rm tail)^2}={4G^3M^2\over 5c^{11}}
\left(\int^{+\infty}_0\! d\tau I^{(5)}_{ij} (U-\tau)
\left[\ln \left({c\tau\over 2r_0}\right)+{11\over 12}\right] \right)^2 \ .
\end{equation}
Finally, there is also the direct 3PN contribution of tails of tails
in (56):

\begin{eqnarray}
 {\cal L}_{\rm tail (\rm tail)}&=&{4G^3M^2\over 5c^{11}}
 I^{(3)}_{ij}(U) \int^{+\infty}_0\! d\tau I^{(6)}_{ij}
 (U-\tau)\nonumber \\ &&\qquad\qquad\times \left[\ln^2
 \left({c\tau\over 2r_0}\right) + {57\over 70} \ln \left({c\tau \over
 2r_0}\right) + {124627\over 44100}\right]\ .
\end{eqnarray}
By a control of all the hereditary integrals in ${\cal L}$ up to 3PN
we have checked \cite{B98tail} that the terms (59)-(60) do exist.  The
two contributions (59) and (60) appear somewhat on the same footing
--~of course both should be taken into account in practical
computations.  Note that in a physical situation where the emission of
radiation stops after a certain date, in the sense that the source
multipole moments become constant after this date (assuming a
consistent matter model which would do this at a given post-Newtonian
order), the only contribution to ${\cal L}$ which survives after the
end of emission is the 3PN tail-square contribution (59).

\section{Post-Newtonian approximation}

In Sections 2 and 3 we have reasoned upon the formal post-Newtonian
expansion ${\overline h^{\mu\nu}}$ of the near-zone field to obtain
the source multipole moments as functionals of the post-Newtonian
pseudo-tensor ${\overline \tau}^{\mu\nu}$. We have also considered in
Sections 4 and 5 the formal expansion $c\to\infty$ of the radiation
field when holding the multipole moments fixed. Clearly missing in
this scheme is an {\it explicit} algorithm for the computation of
${\overline h}^{\mu\nu}$ in the near zone. No such algorithm (say, in
the spirit of the post-Minkowskian algorithm in Section 4) is known
presently, but a lot is known on the first few post-Newtonian
iterations
\cite{LD17,EIH,Fock,C65,CN69,CE70,Ehl77,Ehl80,Ker80,Ker80',%
Capo81,BRu81,BRu82,AD75,PapaL81,S85,S86,BD89,B95}.

The main difficulty in setting up a post-Newtonian algorithm is the
appearance at some post-Newtonian order of divergent Poisson-like
integrals. This comes from the fact that the post-Newtonian expansion
is actually a near-zone expansion \cite{Fock}, which is valid only in
the region where $r=O(\lambda/c)$, and that such an expansion blows up
when taking formally the limit $r\to +\infty$. For instance, Rendall
\cite{Rend92} has shown that the post-Newtonian solution cannot be
asymptotically flat starting at the 2PN or 3PN level, depending on the
gauge.  This is clear from the structure of the exterior near-zone
expansion (15), which involves many positive powers of the radial
distance $r$.  Thus, one is not allowed in general to consider the
limit $r\to +\infty$. In consequence, using the Poisson integral for
solving a Poisson equation with non-compact-support source at a given
post-Newtonian order is {\it a priori} meaningless. Indeed the Poisson
integral not only extends over the near-zone but also over the regions
at infinity. This means that the Poisson integral does not constitute
the correct solution of the Poisson equation in this context. However,
to the lowest post-Newtonian orders it works; for instance it was
shown by Kerlick \cite{Ker80,Ker80'} and Caporali \cite{Capo81} that
the post-Newtonian iteration (including the suggestion by Ehlers
\cite{Ehl77,Ehl80} of an improvement with respect to previous work
\cite{AD75}) is well-defined up to the 2.5PN order where radiation
reaction terms appear, but that some divergent integrals show up at
the 3PN order.
 
Another difficulty is that the post-Newtonian approximation is in a
sense not self-supporting, because it necessitates information coming
from outside its own domain of validity. Of course we have in mind the
boundary conditions at infinity which determine the radiation reaction
in the source's local equations of motion. Again, to the lowest
post-Newtonian orders one can circumvent this difficulty by
considering {\it retarded} integrals that are formally expanded when
$c\to\infty$ as series of ``instantaneous'' Poisson-like integrals
\cite{AD75}. However, this procedure becomes incorrect at the 4PN
order, not to mention the problem of divergencies, because the
near-zone field (as well as the source's dynamics) ceases to be given
by an instantaneous functional of the source parameters, due to the
appearance of ``tail-transported'' hereditary integrals modifying the
lowest-order radiation reaction damping \cite{BD88,B97}.

Let us advocate here that the cure of the latter difficulty (and
perhaps of all difficulties) is the matching equation (16).  Indeed
suppose that one knows a particular solution of the Poisson equation
at some post-Newtonian order. This solution might be in the form of
some ``finite part'' of a Poisson integral. The correct post-Newtonian
solution will be the sum of this particular solution and of a
homogeneous solution satisfying the Laplace equation, namely a
harmonic solution, regular at the origin, which can always be written
in the form $\sum A_L {\hat x}_L$, for some unknown constant tensors
$A_L$. The homogeneous solution is associated with radiation reaction
effects. Now the matching equation states that the multipole expansion
of the post-Newtonian solution agrees with the near-zone expansion of
the exterior field (which has been computed beforehand in Section
4). The multipole expansion of the known particular solution can be
obtained by a standard method, and the multipole expansion of the
homogeneous solution is simply itself, i.e. ${\cal M}(\sum A_L {\hat
x}_L)=\sum A_L {\hat x}_L$. Therefore, we see that the matching
equation determines in principle the homogeneous solution (i.e. all
the unknown tensors $A_L$), and since the exterior field satisfies
relevant boundary conditions at infinity, the $A_L$'s should
correspond to the radiation reaction on a truly isolated system. See
\cite{BD84,BD88,B93,B97} for implementation of this method to
determine the radiation reaction force to 4PN order (1.5PN relative
order).

\subsection{The inner metric to 2.5PN order}

Going to high post-Newtonian orders can become prohibitive because of
the rapid proliferation of terms. Typically any allowed term
(compatible dimension, correct index structure) does appear with a
definite non-zero coefficient in front. However, high post-Newtonian
orders can be manageable if one chooses some appropriate matter
variables, and if one avoids expanding systematically the retardations
due to the speed of propagation of gravity. Often it is sufficient,
and clearer, to present a result in terms of matter variables still
containing some $c$'s, and perhaps also in terms of some convenient
retarded potentials (being clear that any retardation going to an
order higher than the prescribed post-Newtonian order of the
calculation is irrelevant). See for instance (65) and (68)-(69)
below. Anyway, only in a final stage, when a result to the prescribed
order is in hands, should we introduce the more basic matter variables
(e.g. the coordinate mass density) and perform all necessary
retardations. Then of course one does not escape to a profusion of
terms, but at least we have been able to carry the post-Newtonian
iteration using some reasonably simple expressions.

The matter variables are chosen \cite{BD89,B95} in a way consistent
with our earlier definitions (35), i.e.
 
\begin{mathletters}
\begin{eqnarray}
 \sigma &\equiv& {T^{00}+T^{ii}\over c^2}\ ; \\ \sigma_i &\equiv&
 {T^{0i}\over c} \ ;\\ \sigma_{ij} &\equiv& T^{ij} \ .
\end{eqnarray}
\end{mathletters}
To 2.5PN order one defines some {\it retarded} potentials $V$, $V_i$,
${\hat W}_{ij}$, ${\hat X}$ and ${\hat R}_i$, with $V$ and $V_i$
looking like some retarded versions of the Newtonian and
gravitomagnetic potentials, and ${\hat W}_{ij}$ being associated with
the matter and gravitational-field stresses:
 
\begin{mathletters}
\begin{eqnarray}
V &\equiv& \Box^{-1}_R\left\{-4\pi G \sigma\right\} \ ,\\ V_i &\equiv&
\Box^{-1}_R\left\{-4\pi G \sigma_i\right\}\ , \\ {\hat W}_{ij}
&\equiv& \Box^{-1}_R\left\{-4 \pi G (\sigma_{ij} - \delta_{ij}
\sigma_{kk}) - \partial_i V \partial_j V\right\} \ ,\\ 
{\hat R}_i &\equiv& \Box^{-1}_R\left\{ - 4\pi G (V\sigma_i - V_i \sigma) - 2 
\partial_k V
\partial_i V_k - {3\over 2} \partial_t V \partial_i V \right\}\ ,\\
{\hat X} &\equiv& \Box^{-1}_R\biggl\{ -4\pi G V \sigma_{ii} 
+ 2 V_i \partial_t \partial_i V +V \partial_t^2 V  \nonumber \\ 
&&\quad\qquad+{3\over 2} (\partial_t V)^2 - 2 \partial_i V_j \partial_j V_i 
+ \hat{W}_{ij} \partial^2_{ij} V \biggr\} \ ,
\end{eqnarray}
\end{mathletters}
where $\Box^{-1}_R$ denotes the retarded integral (11). All these
potentials but $V$ and $V_i$ have a spatially non-compact support. The
highest non-linearity entering them is cubic; it appears in the last
term of ${\hat X}$.

Based on the latter potentials one can show \cite{B95,BFP98} that the
inner metric to order 2.5PN (in harmonic coordinates,
$\partial_\nu(\sqrt{-g}g^{\mu\nu})=0$) takes the form

\begin{mathletters}
\begin{eqnarray}
g_{00} &=& -1 + {2\over c^2}V - {2\over c^4}V^2 + {8\over c^6}
\left[\hat{X} + V_iV_i + {V^3\over 6}\right] +O\left({1\over
c^8}\right)\ ,\\ g_{0i} &=& -{4\over c^3} V_i - {8\over c^5} \hat{R}_i
+ O\left({1\over c^7}\right)\ ,\\ g_{ij} &=& \delta_{ij} \left( 1 +
{2\over c^2} V + {2\over c^4} V^2
\right) + {4\over c^4} \hat{W}_{ij} + O\left({1\over c^6}\right)\ ,
\end{eqnarray}
\end{mathletters}
(writing ${\overline g}_{\mu\nu}$ would be more consistent with the
notation of Section 2). With this form, we believe, the computational
problems encountered in applications are conveniently divided into the
specific problems associated with the computation of the various
potentials (62), which constitute in this approach some appropriate
computational ``blocks'' (having of course no physical signification
separately). By expanding all powers of $1/c$ present into the matter
densities (61) and into the retardations of the potentials (62), we
find that the metric (63) becomes extremely complicated, as it really
is (see e.g. \cite{CN69,CE70,Ker80,Ker80'}).

Because of our use of retarded potentials, the metric (63) involves
explicitly only even post-Newtonian terms (using the post-Newtonian
terminology that even terms correspond to even powers of $1/c$ in the
equations of motion). We have checked \cite{B95} that the {\it odd}
post-Newtonian terms (responsible for radiation reaction), contained
in (63) via the expansion of retardations, match, in the sense of the
equation (16), to the exterior metric satisfying the no-incoming
radiation condition (9).

The harmonic gauge condition implies some differential equations to be
satisfied by the previous potentials. To 2.5PN order we find

\begin{mathletters}
\begin{eqnarray}
&& \partial_t\left\{V+{1\over c^2}\left[{1\over
2}\hat{W}_{ii}+2V^2\right]\right\} +\partial_i\left\{V_i+{2\over c^2}
\left[\hat{R}_i+VV_i\right]\right\}= O\left({1\over c^4}\right)\ , 
\qquad  \\ 
&& \partial_t V_i +\partial_j\left\{{\hat W}_{ij}-{1\over
2}\delta_{ij} {\hat W}_{kk} \right\}=O\left({1\over c^2}\right)\ ,
\end{eqnarray}
\end{mathletters}
where ${\hat W}_{ii}\equiv\delta_{ij}{\hat W}_{ij}$. These equations
are in turn equivalent to the equation of continuity and the equation
of motion for the matter system,

\begin{mathletters}
\begin{eqnarray}
\partial_t\sigma+\partial_i\sigma_i &=& {1\over c^2}
\left(\partial_t\sigma_{ii}-\sigma\partial_tV\right)+O\left({1\over c^4}
\right)\ , \\
\partial_t\sigma_i+\partial_j\sigma_{ij} &=& \sigma\partial_iV
+O\left({1\over c^2}\right)\ .
\end{eqnarray}
\end{mathletters}
Note that the precision is 1PN for the equation of continuity but only
Newtonian for the equation of motion.

\subsection{The mass-type source moment to 2.5PN order}

From the 2.5PN metric (63) we obtain the pseudo-tensor ${\overline
\tau}$ and the auxiliary quantities (35), that we replace into the
formulas (36) to obtain the 2.5PN source multipole moments. Recall
that the $z$-integration in the moments is to be carried out using the
formula (27). Let us first see how this works at the 1PN order.

We need ${\Sigma}$ to 1PN order and ${\Sigma}_i$ to Newtonian
order. The latter quantity reduces to the matter part,
${\Sigma}_i=\sigma_i+O(1/c^2)$, and the former one reads after a
simple transformation

\begin{equation}
 {\Sigma}=\sigma-{1\over 2\pi Gc^2}\Delta (V^2) +O\left({1\over
 c^4}\right)\ .
\end{equation}
The substitution into the moments $I_L$ given by (36a) leads to

\begin{eqnarray}
I_L&=&\hbox{FP}_{B=0} \int d^3{\bf x}~|{\bf x}/r_0|^B\biggl\{\hat
x_L\sigma -{\hat x_L\over 2\pi Gc^2}\Delta (V^2)\nonumber\\
&&\qquad\quad +{|{\bf x}|^2\hat x_L\over 2c^2(2l+3)}
\partial_t^2\sigma - {4(2l+1)\hat x_{iL}\over
  c^2(l+1)(2l+3)}\partial_t\sigma_i \biggr\}
+O\left({1\over c^4}\right)\ . \qquad
\end{eqnarray}
The integrand is non-compact-supported because of the contribution of
the second term, and accordingly we keep the regularization factor
$|{\bf x}/r_0|^B$ and finite part operation. But let us operate by
parts the second term, using the fact that $|{\bf x}|^B{\hat
x_L}\Delta (V^2) -\Delta(|{\bf x}|^B{\hat x_L})V^2=\partial_i\{|{\bf
x}|^B{\hat x_L}\partial_i(V^2)-\partial_i(|{\bf x}|^B{\hat x_L})V^2\}$
is a pure divergence. When the real part of $B$ is a large {\it
negative} number, we see thanks to the Gauss theorem that the latter
divergence will not contribute to the moment, therefore by the unicity
of the analytic continuation it will always yield zero
contribution. Thus, using $\Delta {\hat x_L}=0$, we can replace $|{\bf
x}|^B{\hat x_L}\Delta (V^2)$ in the second term of (67) by
$\Delta(|{\bf x}|^B{\hat x_L})V^2=B(B+l+1)|{\bf x}|^{B-2}{\hat
x_L}V^2$, and because of the explicit factor $B$ we see that the
second term can be non-zero only in the case where the factor $B$
multiplies an integral owning a simple pole $\sim 1/B$ due to the
integration bound $|{\bf x}|\to\infty$. Expressing $V^2$ (to Newtonian
order) in terms of source points ${\bf z}_1$ and ${\bf z}_2$, we
obtain the integral $\int d^3{\bf x}~|{\bf x}|^{B-2}{\hat x_L} |{\bf
x}-{\bf z}_1|^{-1}|{\bf x}-{\bf z}_2|^{-1}$. When $|{\bf x}|\to\infty$
each $|{\bf x}-{\bf z}_{1,2}|^{-1}$ can be expanded as a series of
${\hat n}_{L_{1,2}} |{\bf x}|^{-l_{1,2}-1}$; then performing the
angular integration shows that the sum of ``multipolarities''
$l+l_1+l_2$ is necessarily an even integer. When this is realized the
remaining radial integral reads $\int d|{\bf x}|~|{\bf
x}|^{B+l-l_1-l_2-2}$ which develops a pole only when
$l-l_1-l_2-2=-1$. But that is incompatible with the previous
finding. Thus the second term in (67) is identically zero, and we end
up simply with a compact-support expression on which we no longer need
to implement the finite part,

\begin{equation}
I_L=\int d^3{\bf x}\biggl\{\hat x_L\sigma 
+{|{\bf x}|^2\hat x_L\over 2c^2(2l+3)}
\partial_t^2\sigma - {4(2l+1)\hat x_{iL}\over
  c^2(l+1)(2l+3)}\partial_t\sigma_i \biggr\} +O\left({1\over
  c^4}\right) .
\end{equation}
This expression was first obtained in \cite{BD89} using a different
method valid at 1PN order. Here we have recovered the same expression
from the formula (36a) valid to any post-Newtonian order
\cite{B95,B98mult}.

Only starting at the 2PN order does the mass multipole moment have a
non-compact support (so the finite part becomes crucial at this
order). By a detailed computation in \cite{B95} we arrive at the
following 2PN (or rather 2.5PN) expression:

\begin{eqnarray}
 I_L(t) &=& {\rm FP}_{B=0}
\int d^3{\bf x}~|{\bf x}/r_0|^B 
 \biggl\{ \hat x_L \biggl[\sigma + {4\over c^4}\sigma_{ii}V \biggr] +
 {|{\bf x}|^2\hat x_L\over 2c^2(2\ell+3)}
 \partial^2_t\sigma\nonumber\\ && +{|{\bf x}|^4\hat x_L\over
 8c^4(2\ell+3)(2\ell+5)} \partial^4_t\sigma - {2(2\ell+1)|{\bf
 x}|^2\hat x_{iL}\over c^4(\ell+1)(2\ell+3)(2\ell+5)}
 \partial^3_t\sigma_i \nonumber\\ && + {2(2\ell+1)\hat x_{ijL}\over
 c^4(\ell+1)(\ell+2)(2\ell+5)} \partial_t^2 \left[ \sigma_{ij} +
 {1\over 4\pi G} \partial_i V \partial_j V \right] \nonumber\\ && +
 {\hat x_L \over \pi Gc^4} \biggl[ -{\hat W}_{ij} \partial_{ij}^2 V -
 2 V_i\partial_t\partial_i V + 2\partial_i V_j \partial_j V_i -
 {3\over 2} (\partial_t V)^2 -V\partial_t^2 V\biggr] \nonumber \\ &&
 -{4(2\ell+1)\hat x_{iL}\over c^2(\ell+1)(2\ell+3)} \partial_t \left[
 \left( 1 +{4V\over c^2} \right) \sigma_i \right. \nonumber\\
 &&\left.+ {1\over \pi Gc^2} \left( \partial_k V[\partial_i V_k
 -\partial_k V_i] + {3\over 4} \partial_t V \partial_i V \right)
 \right] \biggr\} +O\left({1\over c^6}\right)\ .
\end{eqnarray}
Recall that the canonical moment $M_L$ differs from the source moment
$I_L$ at precisely the 2.5PN order [see (46)].

\section{Point-particles}

So far the post-Newtonian formalism has been developed for {\it
smooth} (i.e. $C^\infty$) matter distributions. As such, the source
multipole moments (36) become ill-defined in the presence of
singularities. We now argue that the formalism is in fact also
applicable to singular sources (notably point-particles described by
Dirac measures) provided that we add to our other basic assumptions a
certain method for removing the infinite self-field of
point-masses. Our main motivation is the inspiralling compact binary
-- a system of two compact objects (neutron stars or black holes)
which can be described with great precision by two point-particles
moving on a circular orbit, and whose orbital phase evolution should
be computed prior to gravitational-wave detection with relative 3PN
precision \cite{3mn,CFPS93}.

For this application we restrict ourselves to two point-masses $m_1$
and $m_2$ (constant Schwarzschild masses). The trajectories are ${\bf
y}_1(t)$ and ${\bf y}_2(t)$ and the coordinate velocities ${\bf
v}_{1,2}=d{\bf y}_{1,2}/dt$; we pose $v_{1,2}^\mu=(c,{\bf
v}_{1,2})$. The symbol $1\leftrightarrow 2$ means the same term but
with the labels of the two particles exchanged.  A model for the
stress-energy tensor of point-masses (say, at 2PN order) is

\begin{mathletters}
\begin{eqnarray}
&& T^{\mu\nu}_{\rm point-mass}({\bf x},t)= \mu_1(t) v_1^\mu(t)
v_1^\nu(t)
\delta[{\bf x}-{\bf y}_1(t)] + 1\leftrightarrow 2 \ ; \\
&& \mu_1(t) \equiv {m_1 \over \sqrt{(gg_{\rho\sigma})_1 
{\displaystyle \frac{v_1^\rho v_1^\sigma}{c^2}}}} \ ,
\end{eqnarray}
\end{mathletters}
where $\delta$ denotes the three-dimensional Dirac measure, and
$g_{\mu\nu}$ the metric coefficients in harmonic coordinates ($g\equiv
{\rm det}g_{\mu\nu}$). The notation $(gg_{\mu\nu})_1$ means the value
at the location of particle 1.  However, due to the presence of the
Dirac measure at particles 1 and 2, the metric coefficients will be
singular at 1 and 2.  Therefore, we must supplement the model (70) by
a method of ``regularization'' able to give a sense to the ill-defined
limit at 1 or 2. {\it A priori} the choice of one or another
regularization constitutes a fully-qualified element of the model of
point-particles. In the following we systematically employ the
Hadamard regularization, based on the Hadamard ``partie finie'' of a
divergent integral \cite{Hadamard,Schwartz}.

Let us discuss an example. The ``Newtonian'' potential $U$, defined by
$U=\Delta^{-1}(-4\pi G \sigma)$, where $\sigma$ is given by (61a) [we
have $V=U+O(1/c^2)$], follows from (70a) as

\begin{equation}
U ={G\mu_1\over r_1} \left[1+{v_1^2\over c^2}\right] +1\leftrightarrow
2\ ,
\end{equation}
where $r_1 =|{\bf x}-{\bf y}_1|$. To Newtonian order
$U=Gm_1/r_1+O(1/c^2)+1\leftrightarrow 2$. We compute $U$ at the 1PN
order: from (70b) we deduce at this order $\mu_1/m_1 =1 -(U)_1/c^2
+v^2_1/2c^2+O(1/c^4)$, which involves $U$ itself taken at point 1, but
of course this does not make sense because $U$ is singular at 1 and
2. Now, after applying the Hadamard regularization (described below),
we obtain unambiguously the standard Newtonian result
$(U)_1=Gm_2/r_{12}+O(1/c^2)$, where $r_{12}=|{\bf y}_1-{\bf y}_2|$,
that we insert back into $\mu_1$. So, $U$ at 1PN, and its regularized
value at 1, read
  
\begin{mathletters}
\begin{eqnarray}
U&=&{Gm_1\over r_{1}}\biggl(1+{1\over c^2}\biggl[-{Gm_2\over r_{12}}
+{3\over 2}v_1^2\biggr]\biggr)+O\left({1\over c^4}\right)
+1\leftrightarrow 2\ , \\ (U)_1&=&{Gm_2\over r_{12}}\biggl(1+{1\over
c^2}\biggl[-{Gm_1\over r_{12}} +{3\over
2}v_2^2\biggr]\biggr)+O\left({1\over c^4}\right)\ .
\end{eqnarray}
\end{mathletters}

\subsection{Hadamard partie finie regularization}

We consider the class of functions of the field point ${\bf x}$ which
are smooth on $I\!\!R^3$ except at the location of the two source
points ${\bf y}_{1,2}$, around which the functions admit some
power-like expansions in the radial distance $r_1=|{\bf x}-{\bf
y}_1|$, with fixed spatial direction ${\bf n}_1=({\bf x}-{\bf
y}_1)/r_1$ (and idem for 2). Thus, for any $F({\bf x})$ in this class,
we have

\begin{mathletters}
\begin{eqnarray}
F&=&\sum_a r_1^a f_{1(a)}({\bf n}_1) \qquad (\hbox{when $r_1\to 0$})\
;\\ F&=&\sum_a r_2^a f_{2(a)}({\bf n}_2) \qquad (\hbox{when $r_2\to
0$})\ ,
\end{eqnarray}
\end{mathletters}
where the summation index $a$ ranges over values in $Z\!\!\!Z$ bounded
from below, $a\geq -a_0$ (we do not need to be more specific), and
where the coefficients of the various powers of $r_{1,2}$ depend on
the spatial directions ${\bf n}_{1,2}$. In (73) we do not write the
remainders for the expansions because we don't need them; simply, we
regard the expansions (73) as listings of the various coefficients
$f_{1(a)}$ and $f_{2(a)}$. We assume also that the functions $F$ in
this class decrease sufficiently rapidly when $|{\bf x}|\to\infty$, so
that all integrals we consider are convergent at infinity.

The integral $\int d^3{\bf x} F$ is in general divergent because of
the singular behaviour of $F$ near ${\bf y}_{1,2}$, but we can compute
its partie finie (${\rm Pf}$) in the sense of Hadamard
\cite{Hadamard,Schwartz}. Let us consider two volumes surrounding the
two singularities, of the form $r_1\leq s \rho_1 ({\bf n}_1)$ (and
similarly for 2), where $s$ measures the size of the volume and
$\rho_1$ gives its shape as a function of the direction ${\bf n}_1$
($\rho_1=1$ in the case of a spherical ball). Using (73) it is easy to
determine the expansion when $s\to 0$ of the integral extending on
$I\!\!R^3$ deprived from the two previous volumes, and then to
subtract from the integral all the divergent terms when $s\to 0$ in
the latter expansion. The Hadamard partie finie is defined to be the
limit when $s\to 0$ of what remains. As it turns out, the result can
be advantageously re-expressed in terms of an integral on $I\!\!R^3$
deprived from two {\it spherical} balls ($\rho_{1,2}=1$), at the price
of introducing two constants $s_{1,2}$ which depend on the shape of
the two regularizing volumes originally considered. With full
generality the Hadamard partie finie of the divergent integral reads

\begin{eqnarray}
&&{\rm Pf}\int d^3{\bf x}~ F\equiv \lim_{s\to
0}\biggl\{\int_{r_1>s\atop r_2>s} d^3{\bf x} \ F\nonumber\\
&&\qquad+\!\!\!\sum_{a+3\leq -1}{s^{a+3}\over a+3}\int d\Omega_1
f_{1(a)} +\ln\left({s\over s_1}\right)\int d\Omega_1 f_{1(-3)}
+1\leftrightarrow 2\biggr\}
\end{eqnarray}
where $s_1$ is given by

\begin{equation}
\ln s_1={\int d\Omega_1f_{1(-3)}\ln\rho_1 \over\int 
d\Omega_1f_{1(-3)}} \ .
\end{equation}
Because of the two arbitrary constants $s_{1,2}$ the Hadamard partie
finie is ambiguous, and one could think {\it a priori} that there is
no point about defining a divergent integral by means of an ambiguous
expression.  Actually the point is that we control the origin of these
constants: they come from the coefficients of $1/r_{1,2}^3$ in the
expansions of $F$, which generate logarithmic terms in the
integral. As we shall see the constants $s_{1,2}$ do not appear in the
post-Newtonian metric up to the 2.5PN order (they are expected to
appear only at 3PN order).

We can also give a meaning to the value of the function $F$ at the
location of particle 1 for instance, by taking the average over all
directions ${\bf n}_1$ of the coefficient of the zeroth power of $r_1$
in (73a), namely

\begin{equation}
(F)_1\equiv \int {d\Omega_1\over 4\pi} f_{1(0)}\ .
\end{equation}
We refer also to the definition (76) as the Hadamard partie finie (of
the function $F$ at 1) because this definition is closely related to
the definition (74) of the Hadamard partie finie of a divergent
integral.  To see this, apply (74) to the case where the function $F$
is actually a gradient, $F=\partial_iG$, where $G$ satisfies (73) [it
is then clear that $F$ itself satisfies (73)]. We find

\begin{equation}
{\rm Pf}\int d^3{\bf x}~ \partial_iG=-4\pi (n_1^i r_1^2G)_1
-4\pi (n_2^i r_2^2G)_2
\end{equation}
where in the right side the values at 1 and 2 are taken in the sense
of the Hadamard partie finie (76). This nice connection between the
Hadamard partie finie of a divergent integral and that of a singular
function is clearly understood from applying the Gauss theorem on two
surfaces $r_{1,2}=s$ surrounding the singularities (there is no
dependence on the constants $s_{1,2}$).

\subsection{Multipole moments of point-mass binaries}

To compute the source moments (36) of two point-particles we insert
(70) in place of the stress-energy tensor $T^{\mu\nu}$ of a continuous
source, and we pick up the Hadamard partie finie [in the sense of
(74)] of all integrals. This {\it ansatz} reads
 
\begin{mathletters}
\begin{eqnarray}
(I_L)_{\rm point-mass} &=& {\rm Pf} \left\{ I_L [T^{\mu\nu}_ {\rm
point-mass}] \right\} \ ;\\ (J_L)_{\rm point-mass} &=& {\rm Pf}
\left\{ J_L [T^{\mu\nu}_ {\rm point-mass}] \right\}\ .
\end{eqnarray}
\end{mathletters}

As we have seen in (69), the source multipole moments involve at high
PN order many (non-compact-support) non-linear contributions which can
be expressed in terms of retarded potentials such as $V$. The paradigm
of such non-linear contributions is a term involving the quadratic
product of two (derivatives of) potentials $V$, say $\partial
V\partial V$, or, neglecting $O(1/c^2)$ corrections, $\partial
U\partial U$.  To Newtonian order $U$ is given by $Gm_1/r_1+Gm_2/r_2$
and it is easily checked that this paradigmatic term can be written as
a certain derivative operator, say $\partial\partial$, acting on the
elementary integral (assuming for simplicity $l=2$)

\begin{equation}
Y_{ij} ({\bf y}_1,{\bf y}_2) \equiv -{1\over 2\pi}{\rm FP}_{B=0}
\int d^3{\bf x}~ |{\bf x}/r_0|^B {\hat{x}_{ij} \over r_1 r_2}\ . 
\end{equation}
We see that the integral would be divergent at infinity without the
finite part operation. However, it is perfectly well-behaved near 1
and 2 where there is no need of a regularization.  The integral (79)
can be evaluated in various ways; the net result is \cite{B95,BDI95}

\begin{equation}
Y_{ij} ={r_{12}\over 3} 
\biggl[y_1^{<ij>}+y_1^{<i}y_2^{j>}+y_2^{<ij>}\biggr]
\end{equation}
where $<ij>\equiv {\rm STF}(ij)$.  Starting at 3PN order we meet some
elementary integrals which need the regularization at 1 or 2 in
addition to involving the finite part at infinity. An example is

\begin{equation}
Z_{ij}({\bf y}_1) \equiv -{1\over 2\pi} {\rm Pf} \biggl\{ {\rm
FP}_{B=0}
\int d^3{\bf x}~ |{\bf x}/r_0|^B {\hat{x}_{ij}\over r^3_1}\biggr\} \ .
\end{equation}
To obtain this integral one splits it into a near-zone integral
extending over the domain $r_1< {\cal R}_1$ (say), and a far-zone
integral extending over ${\cal R}_1 <r_1$. The Hadamard regularization
at 1 applies only to the near-zone integral, while the finite part at
$B=0$ is needed only for the far-zone integral. The result, found to
be independent of the radius ${\cal R}_1$, reads \cite{BIJ99}

\begin{equation}
Z_{ij} = \biggl[2\ln\left(s_1\over r_0\right)+{16\over 15}\biggr]\,
y_1^{<ij>} \ .
\end{equation}
In this case we find an explicit dependence on both the constants
$r_0$ due to the finite part at infinity, and $s_1$ due to the
Hadamard partie finie near 1 [see (74)]. However these constants do
not enter the multipole moments before the 3PN order (collaboration
with Iyer and Joguet \cite{BIJ99}).

A long computation, done in \cite{BDI95}, yields the mass-type
quadrupole moment at the 2PN order fully reduced in the case of two
point-masses moving on a circular orbit.  The method is to start from
(69) (issued from \cite{B95}) and to employ notably the elementary
integral (79)-(80) (see also \cite{BDI95} for the treatment of a
cubically non-linear term). An equivalent result has been obtained by
Will and Wiseman using their formalism \cite{WWi96}.  In a
mass-centered frame the moment is of the form

\begin{equation}
I_{ij} = \mu \biggl(A\,{\hat y}_{ij} + B\,{{\hat
v}_{ij}\over\omega^2}\,\biggr)+O\left({1\over c^5}\right)
\ ,
\end{equation}
where $y_i=y_1^i-y_2^i$ and $v_i=v_1^i-v_2^i$, where $\omega$ denotes
the binary's Newtonian orbital frequency [$\omega^2=Gm/r_{12}^3$ with
$m=m_1+m_2$], and where $\mu=m_1m_2/m$ is the reduced mass. The point
is to obtain the coefficients $A$ and $B$ developed to 2PN order in
terms of the post-Newtonian parameter $\gamma=Gm/r_{12}c^2$, where we
recall that $r_{12}$ is the distance between the two particles in
harmonic coordinates. Untill 2PN we find some definite polynomials in
the mass ratio $\nu=\mu/m$ (such that $0<\nu\leq 1/4$):

\begin{mathletters}
\begin{eqnarray}
A &=& 1 + \gamma \biggl[-{1\over 42}-{13\over 14}\nu \biggr] 
+ \gamma^2
\biggl[-{461\over 1512} -{18395\over 1512}\nu - {241\over 1512} 
\nu^2\biggr] \ ,\\
B &=& \gamma \biggl[{11\over 21}-{11\over 7}\nu\biggr] + \gamma^2
\biggl[{1607\over 378}-{1681\over 378}
\nu +{229\over 378}\nu^2\biggr] \ .
\end{eqnarray}
\end{mathletters}

The 2PN mass quadrupole moment (83)-(84) is part of a program aiming
at computing the orbital phase evolution of inspiralling compact
binaries to high post-Newtonian order (see Section 7.4). First-order
black-hole perturbations, valid in the test-mass limit $\nu\to 0$ for
one body, have already achieved the very high 5.5PN order
\cite{P93,Sasa94,TSasa94,TTS96}. Recovering the result of black-hole
perturbations in this limit constitutes an important check of the
overall formalism. For the moment it passed the check to 2.5PN order
\cite{BDI95,B96}; this is quite satisfactory regarding the many
differences between the present approach and the black-hole
perturbation method.

\subsection{Equations of motion of compact binaries}

The equations of motion of two point-masses play a crucial role in
accounting for the observed dynamics of the binary pulsar PSR1913+16
\cite{TFMc79,TW82,T93,D83b}, and constitute an important part of the
program concerning inspiralling compact binaries.  The motivation for
investigating rigorously the equations of motion came in part from the
salubrious criticizing remarks of J\"urgen Ehlers {\it et al}
\cite{EhlRGH}. Four different approaches have succeeded in obtaining
the equations of motion of point-mass binaries complete up to the
2.5PN order (dominant order of radiation reaction): the
``post-Minkowskian'' approach of Damour, Deruelle and colleagues
\cite{BeDD81,DD81a,D82,D83a}; the ``Hamiltonian'' approach of
Sch\"afer and predecessors \cite{O74a,O74b,S85,S86} ; the
``extended-body'' approach of Kopejkin {\it et al}
\cite{Kop85,GKop86}; and the ``post-Newtonian'' approach of Blanchet,
Faye and Ponsot \cite{BFP98}. The four approaches yield mutually
agreeing results.

The post-Newtonian approach \cite{BFP98} consists of (i) inserting the
point-mass stress-energy tensor (70) into the 2.5PN metric in harmonic
coordinates given by (63); (ii) curing systematically the self-field
divergences of point-masses using the Hadamard regularization; and
(iii) substituting the regularized metric into the standard geodesic
equations. For convenience we write the geodesic equation of the
particle 1 in the Newtonian-like form

\begin{equation}
{d{\cal P}_1^i\over dt} = {\cal F}_1^i  
\end{equation}
where the (specific) linear momentum ${\cal P}_1^i$ and force ${\cal
F}_1^i$ are given by

\begin{equation}
{\cal P}_1^i =c\left({v_1^\mu g_{i\mu}\over \sqrt{-g_{\rho\sigma}
{\displaystyle v_1^\rho v_1^\sigma}}}\right)_1~; \quad\qquad\quad\quad
{\cal F}_1^i ={c\over 2} \left({v_1^\mu v_1^\nu \partial_i
g_{\mu\nu}\over
\sqrt{-g_{\rho\sigma} {\displaystyle v_1^\rho v_1^\sigma}}}\right)_1 \ .
\end{equation}
Crucial in this method, the quantities are evaluated at the location
of particle 1 according to the rule (76).  All the potentials (62) and
their gradients are evaluated in a way similar to our computation of
$U$ in (72), and then inserted into (85)-(86).  We ``order-reduce''
the result, i.e. we replace each acceleration, consistently with the
approximation, by its equivalent in terms of the positions and
velocities as given by the (lower-order) equations of motion. After
simplication we find, in agreement with other methods,

\begin{eqnarray}
{dv_1^i\over dt} = &-& {Gm_2\over r_{12}^2} n_{12}^i + {Gm_2\over
r_{12}^2c^2}\biggl\{v_{12}^i \left[4(n_{12}v_1) - 3(n_{12}v_2)\right]
\nonumber\\ &+&n_{12}^i \left[ -v^2_1 - 2v^2_2 + 4(v_1v_2) + {3\over
2} (n_{12}v_2)^2 + 5{Gm_1\over r_{12}} + 4{Gm_2\over r_{12}}
\right]\biggr\}\nonumber\\
  &+&{Gm_2\over r_{12}^2c^4} n_{12}^i \biggl\{\left[ -2v^4_2 + 4v^2_2
  (v_1v_2) - 2(v_1v_2)^2 \right.\nonumber \\ &+& \left. {3\over 2}
  v^2_1 (n_{12}v_2)^2 +{9\over 2} v^2_2 (n_{12}v_2)^2 -6(v_1v_2)
  (n_{12}v_2)^2 - {15\over 8} (n_{12}v_2)^4 \right] \nonumber \\
  &+&{Gm_1\over r_{12}}\left[ -{15\over 4} v^2_1 +{5\over 4} v^2_2
  -{5\over 2} (v_1v_2) \right.\nonumber\\ &+&\left. {39\over 2}
  (n_{12}v_1)^2 -39(n_{12}v_1)(n_{12}v_2)+{17\over 2}(n_{12}v_2)^2
  \right]\nonumber\\ &+&{Gm_2\over r_{12}}\left[ 4 v^2_2 - 8(v_1v_2)+
  2(n_{12}v_1)^2 - 4(n_{12}v_1)(n_{12}v_2) -
  6(n_{12}v_2)^2\right]\nonumber \\ &+&{G^2\over r_{12}^2}\left[
  -{57\over 4}m^2_1 - 9m^2_2 - {69\over 2} m_1m_2
  \right]\biggr\}\nonumber\\ &+& {Gm_2\over r_{12}^2c^4}v_{12}^i
  \biggl\{ v^2_1(n_{12}v_2)+4v^2_2(n_{12}v_1) -5v^2_2(n_{12}v_2)
  -4(v_1v_2)(n_{12}v_1)\nonumber\\ &+& 4(v_1v_2)(n_{12}v_2)
  -6(n_{12}v_1)(n_{12}v_2)^2 + {9\over 2} (n_{12}v_2)^3 \nonumber\\
  &+&{Gm_1\over r_{12}} \left[ -{63\over 4}(n_{12}v_1) + {55\over 4}
  (n_{12}v_2)\right] +{Gm_2\over
  r_{12}}\left[-2(n_{12}v_1)-2(n_{12}v_2)\right]\biggr\} \nonumber\\
  &+&{4G^2m_1m_2\over 5c^5r_{12}^3}\biggr\{ n_{12}^i (n_{12}v_{12})
\left[-6{Gm_1\over r_{12}}+{52\over 3}{Gm_2\over 
r_{12}}+3v_{12}^2\right]\nonumber\\
&+&v_{12}^i \left[2{Gm_1\over r_{12}}-8{Gm_2\over r_{12}}-
v_{12}^2\right]\biggl\}+O\left({1\over c^6}\right)\ , 
\end{eqnarray}
[where $n_{12}^i=(y_1^i-y_2^i)/r_{12}$; $v_{12}^i=v_1^i-v_2^i$; and
e.g. $(n_{12}v_{1})$ denotes the Euclidean scalar product]. At the 1PN
or $1/c^2$ level the equations were obtained before by Lorentz an
Droste \cite{LD17}, and by Einstein, Infeld and Hoffmann
\cite{EIH}. The 2.5PN or $1/c^5$ term represents the radiation damping
in harmonic coordinates [correct because the metric (63) we started
with matches to the post-Minkowskian exterior field]. In the case of
circular orbits, the equations simplify drastically:

\begin{equation}
 {dv_{12}^i\over dt} = -\omega^2_{\rm 2PN} y_{12}^i -{32G^3m^3\nu\over
 5c^5r_{12}^4}v_{12}^i + O\left({1\over c^6}\right)\ ,
\end{equation}
where the orbital frequency $\omega_{\rm 2PN}$ of the 2PN circular
motion reads

\begin{equation}
 \omega^2_{\rm 2PN}={Gm\over r_{12}^3} \left[ 1+(-3+\nu) \gamma +
 \left( 6 +{41\over 4} \nu +\nu^2 \right) \gamma^2 \right]
\end{equation}
(the post-Newtonian parameter is $\gamma=Gm/c^2r_{12}$; and
$\nu=\mu/m$).

\subsection{Gravitational waveforms of inspiralling compact binaries} 

The gravitational radiation field and associated energy flux are given
by (52) and (57) in terms of time-derivatives of the radiative
multipole moments, themselves related to the source multipole moments
by formulas such as (56). Furthermore, at a given post-Newtonian
order, the source moments admit some explicit though complicated
expressions such as (68)-(69), which, when specialized to
(non-spinning) point-mass circular binaries, yield e.g. (83)-(84).

Now, for insertion into the radiation field and energy flux, one must
compute the {\it time-derivatives} of the binary moments, with
appropriate order-reduction using the binary's equations of motion
(87)-(89). This yields in particular the fully reduced (up to the
prescribed post-Newtonian order) gravitational waveform of the binary,
or more precisely the two independent ``plus'' and ``cross''
polarization states $h_+$ and $h_\times$. The result to 2PN order is
written in the form

\begin{equation}
   h_{+,\times} = \frac{2Gm\nu x}{c^2 R} \left\{ H^{(0)}_{+,\times} +
   x^{1/2} H^{(1/2)}_{+,\times} + x H^{(1)}_{+,\times} + x^{3/2}
   H^{(3/2)}_{+,\times} + x^2 H^{(2)}_{+,\times} \right\}\ ,
   \label{90}
\end{equation}
where, for convenience, we have introduced a post-Newtonian parameter
which is directly related to the orbital frequency: $x=(Gm\omega_{\rm
2PN}/c^3)^{2/3}$, where $\omega_{\rm 2PN}$ is given for circular
orbits by (89). The various post-Newtonian coefficients in (90) depend
on the cosine and sine of the ``inclination'' angle between the
detector's direction and the normal to the orbital plane ($c_i = \cos
i$ and $s_i = \sin i$), and on the masses through the ratios
$\nu=\mu/m$ and $\delta m/m$, where $\delta m=m_1-m_2$. The result for
the ``plus'' polarization (collaboration with Iyer, Will and Wiseman
\cite{BIWW96}) is

\begin{mathletters}
\begin{eqnarray}
 H^{(0)}_+ &=& -(1+c^2_i) \cos 2\psi \ , \\ H^{(1/2)}_+ &=& -{s_i\over
 8} {{\delta m} \over m} \biggl[ (5+c^2_i) \cos \psi - 9 (1+c^2_i)
 \cos 3\psi \biggr]\ , \\ H^{(1)}_+ &=& {1 \over 6} \biggl[ 19 + 9
 c^2_i - 2 c^4_i - \nu( 19 - 11 c^2_i - 6 c^4_i ) \biggr] \cos 2\psi
 \nonumber \\ &&-{4\over 3} s^2_i (1+c^2_i) (1-3\nu) \cos 4\psi\ , \\
 H^{(3/2)}_+ &=& {s_i \over 192}{{\delta m}\over m} \biggl\{\biggl[ 57
 + 60 c^2_i - c^4_i - 2 \nu (49 - 12 c^2_i - c^4_i) \biggr] \cos \psi
 \nonumber\\ &&- {27 \over 2} \biggl[ 73 + 40 c^2_i - 9 c^4_i - 2 \nu
 (25 - 8 c^2_i - 9 c^4_i) \biggr] \cos 3\psi \cr && + {625\over 2}
 (1-2\nu) s_i^2 (1+c^2_i)\cos 5\psi \biggr\} - 2 \pi (1+c^2_i) \cos
 2\psi \ , \\ H^{(2)}_+ &=& {1 \over 120} \biggl[ 22 + 396 c^2_i + 145
 c^4_i - 5 c^6_i + {5 \over 3} \nu ( 706 - 216 c^2_i - 251 c^4_i + 15
 c^6_i) \nonumber\\ &&\qquad\quad -5 \nu^2(98 - 108 c^2_i + 7 c^4_i +
 5 c^6_i) \biggr] \cos 2 \psi \nonumber\\ &&+ {2 \over 15} s^2_i
 \biggl[ 59 + 35 c^2_i - 8 c^4_i - {5 \over 3} \nu ( 131 + 59 c^2_i -
 24 c^4_i) \nonumber\\ &&\qquad\quad + 5 \nu^2 (21 - 3 c^2_i - 8
 c^4_i) \biggr] \cos 4 \psi \nonumber\\ &&- {81\over 40} (1-5\nu
 +5\nu^2) s^4_i (1+ c^2_i) \cos 6\psi \nonumber\\ &&+ {s_i \over 40}
 {{\delta m} \over m}\biggl\{ \biggl[ 11 + 7 c^2_i + 10 (5+c^2_i) \ln
 2 \biggr] \sin \psi - {5\pi} (5+c^2_i) \cos \psi \nonumber\\ &&
 \qquad\quad - 27 \biggl[ 7 - 10\, \ln (3/2) \biggr] (1+c^2_i) \sin
 3\psi + 135 \pi (1+c^2_i) \cos 3\psi \biggr\}\ .  \nonumber\\
\end{eqnarray}
\end{mathletters}
The ``cross'' polarization admits a similar expression (see
\cite{BIWW96}). Here, $\psi$ denotes a particular phase variable, 
related to the actual binary's orbital phase $\phi$ and frequency
$\omega \equiv \omega_{2PN}$ by

\begin{equation}
 \psi =\phi - {2Gm \omega \over c^3} \ln \left ({\omega \over \omega_0}
\right) \ ; \label{92}
\end{equation}
$\phi$ is the angle, oriented in the sense of the motion, between the
vector separation of the two bodies and a fixed direction in the
orbital plane (since the bodies are not spinning, the orbital motion
takes place in a plane). In (92), $\omega_0$ denotes some constant
frequency, for instance the orbital frequency when the signal enters
the detector's frequency bandwidth; see \cite{BIWW96} for discussion.

The previous formulas give the waveform of point-mass binaries
whenever the frequency and phase of the orbital motion take the values
$\omega$ and $\phi$. To get the waveform as a function of time, we
must replace $\omega$ and $\phi$ by their explicit time evolutions
$\omega (t)$ and $\phi (t)$. Actually, the frequency is the
time-derivative of the phase: $\omega = d\phi /dt$. The evolution of
the phase is entirely determined, for circular orbits, by the energy
balance equation $dE/dt = -{\cal L}$ relating the binding energy $E$
of the binary in the center of mass to the emitted energy flux ${\cal
L}$. $E$ is computed using the equations of motion (87), and ${\cal
L}$ follows from (57) and application of the previous formalism
[changing the radiative moments to the source moments, applying
(83)-(84), etc...]; the net result for the 2.5PN orbital phase
\cite{BDI95,WWi96,B96} is

\begin{eqnarray}
 \phi &=& \phi_0 -{1\over \nu} \biggl\{ \Theta^{5/8} +\left({3715\over
 8064} + {55\over 96}\nu \right) \Theta^{3/8} - {3\over 4}\pi
 \Theta^{1/4} \nonumber \\ && \qquad\qquad + \left( {9275495\over
 14450688} + {284875\over 258048} \nu + {1855\over 2048} \nu^2 \right)
 \Theta^{1/8} \nonumber\\ && \qquad\qquad +\left( -{38645\over 172032}
 - {15\over 2048} \nu\right) \pi \ln \Theta \biggr\} \ ,\label{93}
\end{eqnarray}
where $\phi_0$ is a constant phase (determined for instance when the
frequency is $\omega_0$), and $\Theta$ the convenient dimensionless
time variable

\begin{equation}
\Theta = {c^3\nu \over 5Gm} (t_c - t)\ , \label{94}
\end{equation}
$t_c$ being the instant of coalescence at which, formally, $\omega
(t)$ tends to infinity (of course, the post-Newtonian method breaks
down before the final coalescence).  All the results are in agreement,
in the limit $\nu \to 0$, with those of black-hole perturbation theory
\cite{P93,Sasa94,TSasa94,TTS96}.

\section{Conclusion}

The formalism reviewed in this article permits investigating in
principle all aspects of the problem of dynamics and
gravitational-wave emission of a {\it slowly-moving} isolated system
(with, say, $v/c\sim 0.3$ at most): the generation of waves, their
propagation in vacuum, the back-reaction onto the system, the
structure of the asymptotic field, and most importantly the relation
between the far-field and the source parameters.  Of course, the
formalism is merely post-Newtonian and never ``exact'', but in
applications to astrophysical objects such as inspiralling compact
binaries this should be sufficient provided that the post-Newtonian
approximation is carried to high order.

Furthermore, there are several places in the formalism where some
results are valid formally to any order of approximation. For
instance, the source multipole moments are related to the {\it
infinite} formal post-Newtonian expansion of the pseudo-tensor [see
(18) or (36)], and the post-Minkowskian iteration of the exterior
field is performed to {\it any} non-linear order [see (43)]. In such a
situation, where an infinite approximate series can be defined, there
is the interesting question of its relation to a corresponding element
in the exact theory. For the moment the only solid work concerns the
post-Minkowskian approximation of the exterior vacuum field, which has
been proved to be asymptotic \cite{DSch90}.  Likewise it is plausible
that the expressions of the source multipole moments could be valid in
the case of exact solutions.

The most important part of the formalism where a general prescription
for how to proceed at {\it any} approximate step is missing, is the
post-Newtonian expansion for the field inside the isolated system. For
instance, though the multipole moments are given in terms of the
formal post-Newtonian expansion of the pseudo-tensor, no general
algorithm for computing {\it explicitly} this post-Newtonian expansion
is known. An interesting task would be to define such an algorithm, in
a manner similar to the post-Minkowskian algorithm in Section 4. In
the author's opinion, the post-Newtonian algorithm should be defined
conjointly with the post-Minkowskian algorithm, and should rely on the
matching equation (16), so as to convey into the post-Newtonian field
the information about the exterior metric.

Note that even if a general method for implementing a complete
approximation series is defined, this method may be unworkable in
practical calculations, because not explicit enough. For instance the
post-Minkowskian series (43) is defined in terms of ``iterated''
retarded integrals, but needs to be suplemented by some formulas, to
be used in applications, for the retarded integral of a multipolar
extended source. In this respect it would be desirable to develop the
formulas generalizing (50)-(51) to any non-linear order. This should
permit in particular the study of the general structure of tails,
tails of tails, and so on.

For the moment the only application of the formalism concerns the
radiation and motion of point-particle binaries. Of course it is
important to keep the formalism as general as possible, and not to
restrict oneself to a particular type of source, but this application
to point-particles offers some interesting questions. Indeed, it seems
that the post-Newtonian approximation used conjointly with a
regularization {\it \`a la} Hadamard works well, and that one is
getting closer and closer to an exact (numerical) solution
corresponding to the dynamics and radiation of two black-holes. So, in
which sense does the post-Newtonian solution (corresponding to
point-masses without horizons) approach a true solution for
black-holes? Does the adopted method of regularizing the self-field
play a crucial role? Is it possible to define a regularization
consistently with the post-Newtonian approximation to all orders?

\section*{Acknowledgments}

The author is especially grateful to Bernd Schmidt for discussions and
for remarks which led to improvement of this article.  Stimulating
discussions with Piotr Crusciel, Thibault Damour and Gerhard Sch\"afer
are also acknowledged. The Max-Planck-Institut f\"ur
Gravitationsphysik (Albert-Einstein-Institut) in Potsdam is thanked
for an invitation during which the writing of the article was begun.

\end{document}